\documentclass[12pt]{iopart}
\usepackage{graphicx,xspace}
\usepackage{iopams}
\newcommand{\seq}{{\cal C}}
\newcommand{\Qs}{Q^{(s)}}
\newcommand{\bd}{R}
\newcommand{\bdw}{r}

\newcommand{\Qsl}{Q_{l}^{(s)}}

\newcommand{\dl}{{\delta_l}}

\newcommand{\dk}{{\delta_k}}

\newcommand{\G}{{\cal G}}

\newcommand{\K}{{\cal K}}

\newcommand{\tP}{{\widetilde P}}
\newcommand{\cP}{{\cal P}}
\newcommand{\cF}{{\cal F}}

\newcommand{\dlp}{{\delta_{l+1}}}
\newcommand{\dlm}{{\delta_{l-1}}}
\begin{document}
\title{$\delta$-exceedance records and random adaptive walks}
\author{Su-Chan Park$^1$ and Joachim Krug$^2$}
\address{$^1$Department of Physics, The Catholic University of Korea, Bucheon 14662, Republic of Korea}
\address{$^2$Institut f\"ur Theoretische Physik, Universit\"at zu K\"oln, K\"oln 50937, Germany}
\vspace{10pt}
\begin{indented}
\item[]\today
\end{indented}
\begin{abstract}
We study a modified record process where the $k$'th record in a series of
independent and identically distributed random variables is defined
recursively through the condition $Y_k > Y_{k-1} - \delta_{k-1}$ with a
deterministic sequence $\delta_k > 0$ called the handicap. For
constant $\delta_k \equiv \delta$ and exponentially distributed random
variables it has been shown in previous work
that the process displays a phase transition as a function of $\delta$
between a normal phase where the mean record value increases
indefinitely and a stationary phase where the mean record value remains
bounded and a finite fraction of all entries are records (Park
\textit{et al} 2015 {\it Phys. Rev.} E \textbf{91} 042707). Here we
explore the behavior for general probability distributions and 
decreasing and increasing sequences $\delta_k$, focusing in particular on the case when $\delta_k$ matches
the typical spacing between subsequent records in the underlying
simple record process without handicap. We find that a continuous
phase transition occurs only in the exponential case, but a novel kind
of first order transition emerges when  $\delta_k$ is
increasing. The problem is partly motivated by the dynamics of
evolutionary adaptation in biological fitness landscapes, where
$\delta_k$ corresponds to the change of the 
deterministic fitness component after
$k$ mutational steps. The results for the record process are used to
compute the mean number of steps that a population performs in such a
landscape before being trapped at a local fitness maximum.      
\end{abstract}
\vspace{2pc}
\noindent{\it Keywords}: record process, extreme value theory,
evolutionary dynamics, epistasis, fitness landscape

\submitto{\jpa}
%

\section{\label{Sec:Intro}Introduction}
\subsection{Record processes} 
The mathematical theory of records is concerned with the statistics of
extremes in a time series of random observations. In the standard
setting, an entry $X_n$ in the series is a (upper) record if it
exceeds all previous entries, i.e. if $X_n > \max\{X_1,
X_2,..,X_{n-1}\}$. When the $X_n$'s are independent and identically
distributed (i.i.d.) random variables, the properties of the sequence
of record times and record values 
have been studied in great detail and are now well understood 
\cite{Gli1978,Arnold1998,Wergen2013}. If the common distribution function 
$F$ of $X_n$ is continuous, the statistics of record times is completely universal
regardless of the choice of $F$. In particular, the number of records
up to time $n$ is asymptotically equal to $\ln(n)$~\cite{Gli1978,Arnold1998,Wergen2013}. 

In many applications of record theory the observations are
subject to uncertainty and the definition of record occurrence needs
to be modified \cite{Edery2013}. Two basic situations are conceivable. On the one hand, 
to make sure that spurious records caused by measurement error are not
counted, one demands that the new record should exceed the old one at
least by an amount $\delta > 0$. On the other hand, to avoid missing
any potential events of interest, one relaxes the record condition and
includes observations in the record sequence that are smaller than the
previous record by at most $\delta$. Both situations have been invoked
to motivate the study of \textit{$\delta$-records} defined by the condition \cite{GLS2007,Gouet2012}
\begin{equation}
\label{delta_records}
X_n > \max\{X_1, X_2, ..., X_{n-1} \} + \delta.
\end{equation}  
Specifically, for $\delta < 0$ the events satisfying
(\ref{delta_records}) are referred to as \textit{near-records} \cite{Balakrishnan2005}. 
An immediate consequence of introducing the parameter $\delta$ is that
the strong universality of the statistics of record times is lost and
replaced by an explicit dependence on the tail properties of the
underlying distribution $F$, similar to other modified record
processes involving discreteness \cite{Vervaat1973,Gouet2005},
rounding effects \cite{Wergen2012} or trends \cite{Ballerini1987,Franke2010}. 

Importantly, according to (\ref{delta_records}) the threshold
that the new record has to exceed is defined in terms of the ``true''
record sequence, the maximum process $M_n \equiv \max\{X_1,X_2,..,X_{n}\}$, which (under
the conditions of measurement uncertainty described above) may not
even be observable. A more faithful representation of the measurement
error scenario, in which the next record occurs conditional on the
previous \textit{observed} record, was introduced twenty years ago by Balakrishnan \textit{et al.} under
the name of \textit{$\delta$-exceedance records} \cite{Balakrishnan1996}.  
They considered the case $\delta > 0$ and derived several results for
the case when the underlying distribution $F$ is of exponential or Gumbel
form. Mainly for reasons of mathematical tractability \cite{Edery2013}, subsequent work
has however focused on the problem of $\delta$-records defined by
(\ref{delta_records}). 

\begin{figure}[t]
\centering
\includegraphics[width=0.7\textwidth]{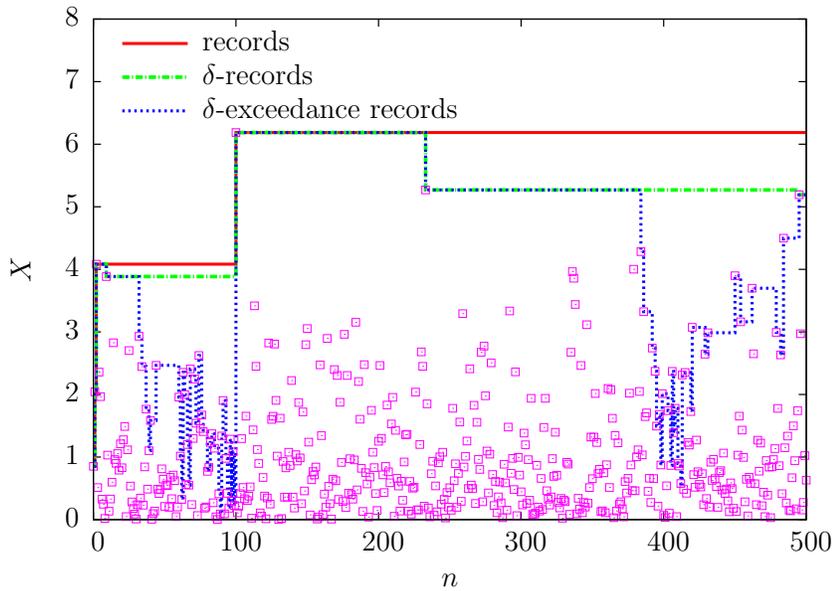} 
\caption{\label{Fig:deltas}  Comparison of sample paths of the standard
  record process (red solid line), $\delta$-records (green dot-dashed line), and
  $\delta$-exceedance records (blue dotted line) for the 
same realization of the background process. Random variables of the background process drawn from an exponential distribution with unit mean are shown as symbols.
In this example, $\delta$ is set to $-1$, 
which corresponds to the critical point
of the $\delta$-exceedance process. Note the cascades of decreasing record values which signal
the incipient stationary phase that emerges when $\delta < -1$. 
}
\end{figure}

A qualitative difference between $\delta$-records and
$\delta$-exceedance records arises in the case when $\delta <
0$; see \fref{Fig:deltas}. Because $\delta$-records are coupled to the 
maximum process $M_n$, the exceedance threshold for a new record defined in
(\ref{delta_records}) increases monotonically in
time. Although the records themselves do not necessarily grow
monotonically, the growing threshold ensures that the record values
increase on average and are pushed into the tail of $F$. By contrast, in the $\delta$-exceedance
record process (to be defined in precise mathematical terms in 
section \ref{Sec:Statement}) the threshold may decrease when $\delta < 0$. This entails the possibility
that a finite fraction of entries in the time series are counted as
records and the expected record value remains asymptotically bounded
even for an unbounded distribution. We will refer to this behavior as
the \textit{stationary} phase of the record process. 

In a recent publication we showed that this scenario is indeed
realized and leads to a novel kind of phase transition as a function
of $\delta$ when the underlying distribution has an exponential tail \cite{PSNK2015}. 
Our results were obtained in the context of adaptive walks in biological
fitness landscapes, which we explain next. 

\subsection{\label{Sec:AW}Adaptive walks}

Adaptive walks are simple evolutionary dynamics defined on a space of
genotypes \cite{KL1987,Macken1989,FL1992,O2002,NK2011,Seetharaman2014a,PSNK2015}. In the most
common setting genotypes are encoded by binary sequences of
length $L$, where each letter denotes the presence of one of two
alleles (say, 0 or 1) at a given genetic locus or nucleotide position. Genotypes are
assigned real fitness values which quantify the reproductive potential
of the corresponding individuals. In one step of the walk,
neighboring genotypes of higher fitness are sampled 
and one of them is chosen as the next position of the walker. Here two
genotypes are defined to be neighbors if they differ by
a single point mutation, that is, at one site of the sequence. If no
fitter neighbors exist the population has reached a
local fitness maximum and the walk stops. Quantities of interest in the theory
of adaptive walks are the number of steps required to reach a local maximum
and the fitness value that has been reached at this point.

In the present work we focus on the conceptually simplest case of
the random adaptive walk (RAW) where the next genotype along the walk is
chosen at random among the neighbors of higher fitness. Let us assume
that the fitness values of different genotypes are i.i.d. random
variables and consider the limit $L \to \infty$. In this limit there are no
local fitness maxima and the walk progresses indefinitely. At each
step the walk moves to a new fitness value which is a random
draw from the underlying fitness distribution conditioned on being
larger than the previous value. Thus, the fitness values
encountered along the walk form a series of records. 

To estimate the number of steps that the RAW takes when $L$ is finite,
we note that the walker will stop when the fitness value that it has
reached exceeds the maximum among $L$ i.i.d. random variables. In the
record picture this means that the time at which the RAW stops is of
order $L$. At this point the number of records, that is, the number of RAW
steps, is of order $\ln(L)$, in agreement with the detailed analysis 
\cite{Macken1989,FL1992}. In fact, the sequence
length $L$ and the time $n$ in the i.i.d. record problem are exactly equivalent in a
variant of the RAW model where the set of available neighboring
genotypes is kept fixed throughout the walk \cite{O2002,NK2011}.  

In \cite{PSNK2015} we considered RAW's in a setting 
where i.i.d. random fitness values are added to a deterministic
fitness profile. The walker is assumed to start in a state of low
fitness and every step brings it closer to a high fitness peak.
Denoting the genotype that has been reached after $k$ steps by
$\seq_k$ and the corresponding fitness by $W(\seq_k)$, the model is
defined by 
\begin{eqnarray}
\label{Wk}
W(\seq_k) = g_k + \eta_{\seq_k},
\end{eqnarray}
where $g_k$ is a deterministic, monotonically increasing function of
$k$ and the $\eta_\seq$'s are i.i.d. random variables associated with
genotypes. The condition for a genotype $\seq'$ to be a possible target for
the next step of the RAW then reads $W(\seq') = g_{k+1} + \eta_{\seq'}
> W(\seq_{k})$ or, in terms of the i.i.d. random variables, 
\begin{eqnarray}
\label{iid}
\eta_{\seq'} > -(g_{k+1} - g_k) + \eta_{\seq_k}.
\end{eqnarray}
For the special case of a linear fitness gradient $g_k = ck$, also
known as the Rough Mount Fuji model \cite{PSNK2015,NSK2014}, the $L \to \infty$ RAW 
is thus seen to be equivalent to the $\delta$-exceedance record
problem with $\delta = - c$. 

\subsection{Goal and outline of the paper}

It was shown in \cite{PSNK2015} that the phase transition in the
$\delta$-exceedance record process with $\delta < 0$ occurs only when
the common distribution of the i.i.d. random variables has an exponential
tail. The special role of the exponential distribution reflects the
well-known fact that record values from exponentially-tailed
distributions are asymptotically equally spaced
\cite{Arnold1998}. Correspondingly, phase-transition like
phenomena can be expected for other tail shapes if the constant
$\delta < 0$ is replaced by an offset that varies with the
number of records, and some preliminary results along these lines were reported
in \cite{PSNK2015}. 

The purpose of this paper is to provide a
comprehensive analysis of the generalized $\delta$-exceedance record problem
with a monotonically varying, negative offset. For notational
convenience, we will denote the offset associated with the $k$'th
record by $-\delta_k$ ($\delta_k > 0$). Moreover, because the offset
facilitates the establishment of new records, $\delta_k$ will be referred to as
the `handicap'. Results covering all three extreme value classes of
random variables and decreasing as well as increasing handicap will
be presented. 

In the context of adaptive walks, the variation of $\delta_k =
g_{k+1}-g_k$ with $k$ implies that the deterministic effect of a
mutation depends on where it occurs along the evolutionary trajectory,
a phenomenon known as epistasis \cite{deVisser2011}. In
particular, a pattern of \textit{diminishing returns epistasis} where
mutational effect sizes decrease with the number
of adaptive steps is commonly observed in evolution
experiments with microbial populations \cite{Chou2011,Khan2011,Wiser2013,Berger2014,Couce2015}. 

In the next section we define the generalized $\delta$-exceedance
record problem considered in this work, and we specify the probability
distributions of the underlying random variables that will be
used. The case of decreasing handicaps is
examined in \sref{Sec:bs} and the case of increasing handicaps
in \sref{Sec:bl}. In \sref{Sec:Walk} 
we use the results for the record process to obtain estimates for
the length of adaptive walks. Finally, in \sref{Sec:Summary} we
summarize our findings and provide some conclusions in the contexts of
record statistics as well as evolutionary dynamics.

\section{\label{Sec:Statement}Statement of the problem}
Consider a sequence $\{X_i, \, i\ge 0\}$ of i.i.d. 
random variables with a common distribution 
function $F$ and probability density $f$. From this sequence, we
construct recursively the generalized $\delta$-exceedance record 
process $\{Y_k, \, k\ge 0\}$ and the corresponding
record-occurrence-time process, $\{n_k, \, k\ge 0\}$, as follows. We first define 
$Y_0 = X_0$ and $n_0 = 0$. Suppose that $Y_l$ and $n_l$ up to $l =
k-1$ ($k=1,2,\ldots,)$ have been determined. We then define $n_k$ as
\begin{eqnarray}
n_k = \mathrm{min}\{i| X_i > Y_{k-1} - \delta_{k-1}, i > n_{k-1} \},
\label{Eq:Process}
\end{eqnarray}
where $\delta_k$ is a deterministic $k$-dependent sequence that will
be called the handicap. Once $n_k$ is determined, we set $Y_k = X_{n_k}$.
For later purposes, $\{X_i, \, i \ge 0\}$ will be referred to as the background process.

In the following, we set 
\begin{eqnarray}
\label{deltak}
\dk = c (k+1)^{b-1} 
\end{eqnarray}
with $b>0$. If $b>1$ ($b<1$), the handicap $\dk$ increases (decreases)
with $k$.
An epistatic fitness landscape model of the form
(\ref{deltak}) has been considered in \cite{Wiehe1997}. 

The following three kinds of distributions
with parameters $a,\bdw,\alpha,\nu,\mu >0$ will be considered for the
background process:
\begin{eqnarray}
F_g(x) =1-\exp(- x^\alpha/a),\nonumber \\
F_w(x) =1-(1- x/\bdw)^{1/\nu},\nonumber\\
F_f(x) =1-(1+ x/a)^{-\mu}.
\label{Eq:distribution}
\end{eqnarray}
The subscripts $g,w,f$ refer to the Gumbel, Weibull, and Fr\'echet
classes of extreme value theory, respectively \cite{deHaan}. 
The corresponding densities $f_g$ and $f_f$ have semi-infinite support $x > 0$ and
the support of $f_w$ is the interval $0 < x < r$. 

As in Ref.~\cite{PSNK2015}, the mean value $z_l$ of the $l$'th record will turn out to 
play an important role in understanding the record process. To derive
a recursion relation for $z_l$ we consider the probability density of $Y_l$, which is denoted by $Q_l$. 
It is a straightforward generalization of the case with
$b=1$~\cite{PSNK2015} to obtain the recursion relation 
\begin{equation}
Q_{l+1}(y) = f(y)\int_{-\infty}^{y+\dl} \frac{Q_{l}(x)}{1 - F(x-\dl)} dx,
\label{Eq:Ql}
\end{equation}
with $Q_0(y)  = f(y)$.
As a consequence, $z_l$ satisfies (see \ref{App:Der} for the derivation)
\begin{eqnarray}
z_{l+1}  = \left \langle \frac{1}{h} \right \rangle_{l+1}
+ \int_\dl^\bd (x-\dl) Q_l(x) dx,
\label{Eq:exactzl}
\end{eqnarray}
where $h$ is the hazard function defined as
\begin{eqnarray}
h(y) \equiv \frac{f(y)}{1-F(y)} = -\frac{d}{dy}\ln \left [  1 - F(y) \right ], 
\end{eqnarray}
$\bd$ is the supremum of the support, 
and $\langle \ldots \rangle_l$ stands for the average with respect to $Q_l$.
The hazard functions corresponding to the distributions~\eref{Eq:distribution} are
\begin{eqnarray}
h_g(x)^{-1} = a x^{1-\alpha}/\alpha,\nonumber\\
h_w(x)^{-1} = \nu(\bdw - x),\nonumber\\
h_f(x)^{-1} = (a+x)/\mu.
\label{Eq:hazard}
\end{eqnarray}
Note that \eref{Eq:exactzl} remains valid for $\bd = \infty$ and
hence can be used for the distributions $F_g$ and $F_f$. Moreover, it can
be shown that the last term in \eref{Eq:exactzl} becomes
$z_l-\dl$ when the support of $f$ is unbounded on both sides and the
recursion reduces to that considered in \cite{PSNK2015}  (see \ref{App:Der}).
The analyses in the following sections will be largely based on \eref{Eq:exactzl}.

Before embarking on the detailed investigation,
it can be instructive to develop a heuristic picture
based on the behavior of the mean record value for the
standard case. It is plausible to expect that the handicaps $\delta_k$ will be
relevant (irrelevant) to the record process if they are asymptotically
larger (smaller) than the mean difference between subsequent record
values in the standard setting. 

Consider first the Gumbel-type
distributions $F_g$, for which the mean value of the $l$'th (standard) 
record is asymptotically equal to $(al)^{1/\alpha}$ \cite{Arnold1998} and hence
the difference between subsequent record values is proportional to
$l^{\frac{1}{\alpha}-1}$. Comparing to (\ref{deltak}) it follows that the
handicap should be relevant (irrelevant) for $b > 1/\alpha$ ($b <
1/\alpha$). The value $b = 1/\alpha$ is thus of special interest as 
it is the only case where an extension of the phase transition
scenario described in \cite{PSNK2015} can be expected to arise. 
For the Fr\'echet class distributions $F_f$
the mean (standard) record value grows exponentially with $l$, whereas it
approaches the upper boundary of the support exponentially fast for the Weibull class distributions
$F_w$ \cite{Arnold1998}. Correspondingly, one expects the record statistics to always be
asymptotically modified (unmodified) by the handicaps for distributions in the Weibull (Fr\'echet)
classes. We will see below that these expectations are largely
confirmed by the detailed analysis, but in addition several unanticipated
features emerge.   
\section{\label{Sec:bs}Decreasing handicap}
When $b = 1$ and $\bd = \infty$, $z_l$ either increases indefinitely with $l$ or 
saturates to a finite value, depending on the tail behavior of $F$~\cite{PSNK2015}. Expecting a similar behavior, 
we focus on determining the conditions under which $z_l$ diverges with
$l$ when $b < 1$.
To this end, we first assume that $z_l$ indeed diverges. Another crucial 
assumption is that $Q_l(x)$ is sharply peaked around $z_l$ when $l$ is 
sufficiently large. Because this amounts to neglecting fluctuations around
$z_l$, it  will be referred to as the mean-field approximation (MFA).
Under these assumptions, we can approximate
the integral in \eref{Eq:exactzl} as
$z_l-\dl$ and, in turn, we arrive at the approximate equation
\begin{eqnarray}
z_{l+1}-z_l \approx \frac{1}{h(z_{l+1})}- \dl.
\label{Eq:dzdla}
\end{eqnarray}
When $\bd$ is finite, the above assumptions are clearly not applicable
and we will use a different approach.

\subsection{\label{Sec:dr_g}Gumbel class}
We first consider the distribution $F_g(x)$. 
Using the corresponding hazard function in \eref{Eq:hazard}, we get
\begin{eqnarray}
z_{l+1}-z_l \approx \frac{a}{\alpha} z_{l+1}^{1-\alpha} - c (l+1)^{b-1}.
\label{Eq:dzdl}
\end{eqnarray}
Assuming that $z_l$ is a slowly varying function of $l$
in the sense that $(z_{l+1}-z_l)/z_l \rightarrow 0$ as $l\rightarrow \infty$, 
we can rewrite the above equation as a differential equation
\begin{eqnarray}
\frac{d}{dl} \left ( \frac{z^\alpha}{a} \right ) 
= 1 - c \frac{\alpha}{a} z^{\alpha-1} l^{b-1},
\label{Eq:lzdiff}
\end{eqnarray}
where $z$ is now meant to be a continuous function of $l$.
Since we are interested in the asymptotic solution,
we set $z^\alpha \approx a (Al)^\gamma$, which yields
\begin{eqnarray}
A^\gamma \gamma l^{\gamma-1} \doteq 1 - A^{1-b}\frac{c\alpha}{a^{1/\alpha}}
\left ( A l \right )^{\gamma(1-1/\alpha) + b-1},
\label{Eq:Agammaz}
\end{eqnarray}
where $\doteq$ means that the equality holds only for the leading behavior
on both sides.
Thus, the asymptotic behavior of $z_l$ can be consistently determined by 
comparing powers, $\gamma-1$ and $\gamma(1-1/\alpha)+b-1$, 
in \eref{Eq:Agammaz} with 0.

If $\gamma(1-1/\alpha)+b-1 >0$, the right hand side (RHS) of \eref{Eq:Agammaz}
will eventually become negative while the left hand side (LHS) 
is positive for any $l$.
Accordingly, any consistent solution requires $\gamma(1-1/\alpha)+b-1\le 0$.
Likewise, if $\gamma>1$, no consistent solution exists because
the LHS increases to infinity while the RHS cannot. 
Hence, we can conclude that only solutions with $\gamma\le 1$ and $\gamma(1-1/\alpha)+b-1 \le 0$ are possible.

Let us first consider what will happen if the solution is $\gamma = 1$.
With this assumption, we rewrite \eref{Eq:Agammaz} as
\begin{eqnarray}
A \doteq 1 - A^{1-b} 
\frac{c\alpha}{a^{1/\alpha}}
\left ( A l \right )^{b-1/\alpha}.
\label{Eq:Agamma1z}
\end{eqnarray}
If $b < 1/\alpha$,  we get $A=1$, which corresponds to the
behavior of the standard record values (that is, $c=0$).
If $b = 1/\alpha$, $A$ is the solution of
equation
\begin{eqnarray}
\label{Eq:Amplitude}
A^{b-1} - A^{b} = \frac{c}{a^{b}b}.
\end{eqnarray}
Note that as long as $b < 1$ a positive solution
of \eref{Eq:Amplitude} uniquely exists in the interval $0< A \le 1$ 
for any $c \ge 0$.
In the following, we will denote this solution by $A_b(c)$.

Now we investigate if a solution with $\gamma < 1$ exists.
Since the LHS of \eref{Eq:Agammaz} approaches 0 in this case, 
the RHS should also approach zero. Thus, we require that
$\gamma = (1-b)/(1-1/\alpha)$ and we get
\begin{eqnarray}
A = \left ( \frac{c\alpha}{a^{1/\alpha}} \right )^{1/(b-1)}
\end{eqnarray}
by equating the right hand size of \eref{Eq:Agammaz} to
zero. Note that for $\gamma$ to be smaller than 1, $b > 1/\alpha$ should be
satisfied. 

Actually, the LHS of \eref{Eq:Agammaz} with
$\gamma <1$ gives rise to a subleading
correction. To see this, let us set 
$z^\alpha= a( A l)^\gamma ( 1 + B l^{-\beta})$,where $A$ and $\gamma$ are 
the solutions for $\gamma<1$ in the above and $\beta>0$. Then we get
\begin{eqnarray}
A^\gamma \gamma l^{\gamma-1} \doteq - \frac{\alpha-1}{\alpha} B l^{-\beta}
\end{eqnarray}
and hence
\begin{eqnarray}
\frac{z_l^\alpha}{a} \approx (A l)^\gamma 
\left [ 1 - \frac{\alpha \gamma}{\alpha-1} A^\gamma l^{\gamma-1}
\right ].
\label{Eq:subzl}
\end{eqnarray}
Since we neglect the effect of fluctuation of $Q_l$, the subleading
term in \eref{Eq:subzl} should however not be taken seriously.

To sum up, \eref{Eq:dzdl} has solutions of diverging $z_l$ for
any $\alpha$ as long as $b<1$. The leading behavior of $z_l$ is
\begin{eqnarray}
\frac{z_l^\alpha}{a}
=(A l)^\gamma,
\label{Eq:zlMFA}
\end{eqnarray}
where
\begin{equation}
\label{Eq:A}
A=\cases{1, &$\alpha < 1/b$,\\
A_b(c), &$\alpha = 1/b$,\\
(c\alpha/a^{1/\alpha})^{1/(b-1)}, & $\alpha > 1/b$,\\}
\end{equation}
and
\begin{equation}
\gamma = \cases{ 
1 ,& $\alpha\le 1/b$ ,\\
(1-b)/(1-1/\alpha), & $\alpha > 1/b$.  \\}
\label{Eq:gamma}
\end{equation}
These results confirm the heuristic considerations of
\sref{Sec:Statement}. For $\alpha < 1/b$ the record values behave
asymptotically as in the standard case, i.e. the handicaps are
irrelevant, whereas for $\alpha > 1/b$ the behavior is modified
qualitatively and the record values grow more slowly than the standard
record process. On the `critical
line' $\alpha = 1/b$ the mean record value grows with the same power
as in the standard case but with a reduced prefactor $A_b(c)$ which
interpolates smoothly between the limits $A_b(c=0) = 1$ and $A_b(c\to\infty) = 0$. 
Thus, in contrast to the case $b=1$, there is no
phase transition as a function of $c$ in the sense of \cite{PSNK2015}. 
For $b=1$ the solution of
\eref{Eq:Amplitude} is $A_1(c)=1-c/a$ which reflects the phase
transition at $c=a$ and is confirmed by the exact solution presented
in \cite{PSNK2015}.

In order to self-consistently check the validity of the MFA  
we need to investigate the behavior of $V_l$, the variance of
$Q_l(x)$. Specifically, the MFA is justified if it can be shown that  
$V_l/z_l^2 \rightarrow 0$ as $l \rightarrow \infty$. The detailed
analysis in \ref{App:Variance} shows that this is indeed the
case. Moreover, although we only used the specific form $1-F_g(x)= e^{-x^\alpha/a}$,
the above conclusions about the leading asymptotic behavior remain valid 
as long as $-\ln [1-F(x)] = -x^\alpha/a + o(x^\alpha)$, that is,
the asymptotic behavior is universal in that the leading behavior
of $-\ln[1-F(x)]$ determines the behavior of $\delta$-exceedance record
values.

\begin{figure}[t]
\centering
\includegraphics[width=0.7\textwidth]{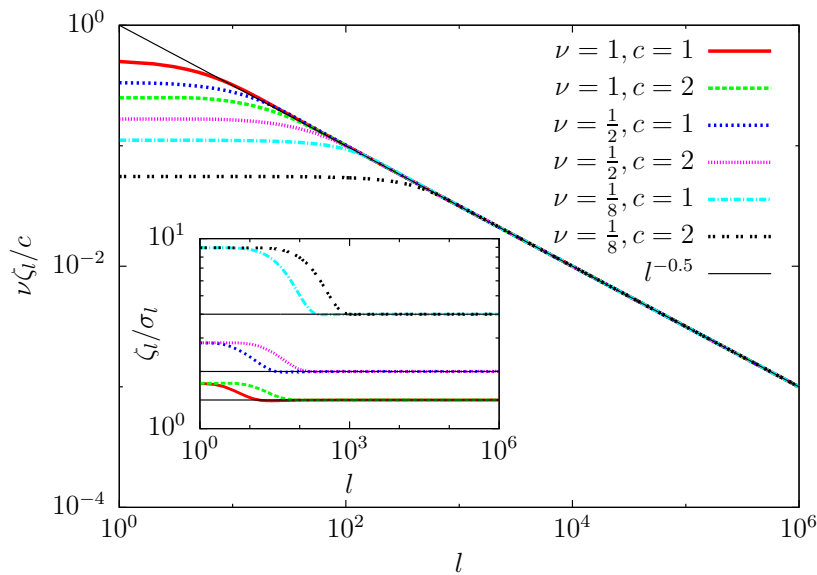}
\caption{\label{Fig:boundz}  Illustration of the approach of the mean record value $z_l$ to the upper boundary $r$ of the support for
Weibull-class distributions. The figure shows $\nu \zeta_l /c$ vs. $l$ for $b=\frac{1}{2}$, $\nu =1$, $\frac{1}{2}$, and $\frac{1}{8}$ with $c=1$ and 2 on a double logarithmic scale. Here $\zeta_l =  r-z_l$. 
For comparison, we plot the predicted behavior $l^{-1/2}$. Inset: Double-logarithmic plots of
$\zeta_l/\sigma_l$ against $l$ for various $\nu$. The horizontal lines 
show the predicted value $\sqrt{2/\nu}$.}
\end{figure}
\subsection{\label{Sec:dr_w}Weibull class}
Now we consider $F_w(x)$. It turns out that it is possible to 
find the generating function 
\begin{eqnarray}
\label{Gcal}
\G_l(\lambda) = \int_0^\bdw e^{\lambda (x-\bdw)} Q_l(x) dx = 
e^{-\bdw\lambda}G_l(-\lambda), 
\end{eqnarray}
where $G_l$ is defined in \eref{Eq:Gdef}. 
From \eref{Eq:Gexact} with $1/h_w(x) = \nu (\bdw-x)$,
we obtain the recursion relation for $\G_l$ in the asymptotic regime as
\begin{eqnarray}
\G_{l+1} = e^{-\lambda \dl} \G_l - \nu \lambda \frac{d\G_{l+1}}{d\lambda},
\label{Eq:Gl_recur}
\end{eqnarray}
where we neglect the contribution of the integral in the domain
$(0,\dl)$. 
If we assume $\G_{l+1} \approx \G_{l}$, \eref{Eq:Gl_recur} becomes
a first order differential equation
\begin{equation}
\nu \lambda \frac{d\G_l}{d\lambda} + \left ( 1 - e^{-\lambda \dl} \right )
\G_l = 0
\end{equation}
with the `initial' condition $\G_l(\lambda = 0)=1$.
Within this approximation scheme,  we get
\begin{eqnarray}
\ln \G_l \approx 
-\frac{1}{\nu}\mathrm{Ein}(\lambda \dl),
\end{eqnarray}
where 
\begin{eqnarray}
\mathrm{Ein}(x)\equiv
\int_0^x \frac{1-e^{-t}}{t} dt
= - \sum_{n=1}^\infty \frac{(-x)^n}{n!n},
\end{eqnarray}
which is an entire function (this function is also found in Ref.~\cite{FL1992}).
Since $\ln \G_l$ is the generating function of cumulants,
the $n$'th cumulant is equal to $\dl^n/(n\nu)$.
For example,
\begin{eqnarray}
\zeta_l\equiv r-z_l = \frac{\dl}{\nu},\qquad
\sigma_l = \frac{\dl}{\sqrt{2\nu}}.
\label{Eq:boundzlsl}
\end{eqnarray}
In \fref{Fig:boundz}, we present simulation results
for $b=\frac{1}{2}$ and for various $\nu$, observing an 
excellent agreement with \eref{Eq:boundzlsl}.

We conclude that in this case 
the approach of the mean record value $z_l$ to the upper boundary $r$ is
completely determined by the behavior of the
handicaps. The approach is algebraic rather than
exponential as in the standard case, confirming our expectation that the
handicaps dominate the record statistics for any $c > 0$ when the
background process belongs to the Weibull class. 

The full distribution of record values 
$Q_l(x)$ can be obtained by inverse Laplace transformation of $\G_l$
such that
\begin{eqnarray}
Q_l(r-x) 
= \frac{1}{2 \pi i} \int_{-i\infty }^{i\infty } 
\exp\left [ \lambda x-\frac{1}{\nu} \mathrm{Ein}(\lambda \dl)\right ] d \lambda
=\frac{1}{\dl} g(x/\dl),
\end{eqnarray}
where 
\begin{eqnarray}
g(z) = \frac{1}{2\pi i } \int_{-i \infty }^{i \infty }
\exp\left [ tz-\frac{1}{\nu} \mathrm{Ein}(t)\right ] dt.
\label{Eq:gy}
\end{eqnarray}
Since $\mathrm{Ein}(t) \sim \ln |t|$ for
$|t| \gg 1$ ($|\mathrm{Arg}(t)| < \pi$),
we find $\G_l(\lambda) \approx (\lambda \dl)^{-1/\nu}$ for
$\lambda \dl \gg 1$,
which implies that $Q_l(r-x)$ behaves as
$x^{1/\nu - 1}$ for small $x$, just like $f_w(x)$.
\subsection{\label{Sec:dr_f}Fr\'echet class}
Now we consider $F_f(x)$ with $1/h_f(x) = (a+x)/\mu$.
Neglecting the contribution from the integral over the domain $(0,\dl)$, 
we get the recursion relation of $z_l$ for large $l$ as
\begin{eqnarray}
\left ( 1 - \frac{1}{\mu}\right ) (z_{l+1}+a) 
\approx (z_l+a) - \dl
\end{eqnarray}
whose solution is
\begin{eqnarray}
\fl
z_l+a =   \left ( \frac{\mu}{\mu-1} \right )^l (z_0 +a)+
\sum_{k=0}^{l-1} \left ( \frac{\mu}{\mu-1} \right )^{l-k} \dk 
\approx \left ( \frac{\mu}{\mu-1} \right )^l (z_0 +a + C_0), 
\end{eqnarray}
where $C_0 = \sum_{k=0}^\infty (1-1/\mu)^k \dk$. 
The above solution is exact when $c = 0$.
Recall that $\mu$ should be larger than 1 in order to have a finite mean $z_l$.
The exponential growth of $z_l$ is identical to the known result
for the i.i.d. record process \cite{Arnold1998}. We conclude that the 
asymptotics is not affected by $\dl$ if $b<1$ and the background
process belongs to the Fr\'echet class.

\section{\label{Sec:bl}Increasing handicap}
This section analyzes the case $b>1$. 
A trivial conclusion for distributions with bounded support such as
$F_w$ is immediate: As soon as $\dl > r$, all 
random variables are records. The effects for unbounded distributions
are more subtle and will be discussed in the following, mostly focusing on
the Gumbel class distributions $F_g$. 

\subsection{\label{Sec:ir_mf}Mean field analysis for the Gumbel class}
As in the previous section,
we first look for a solution with 
diverging $z_l$, assuming $Q_l(x) \approx \delta(x-z_l)$. 
The recursion relation for $z_l$ under the MFA is
\begin{eqnarray}
z_{l+1} - z_l = \frac{a}{\alpha} z_{l+1}^{1-\alpha} - \dl
+ (\dl - z_l) \Theta( \dl-z_l),
\end{eqnarray}
where $\Theta$ is the Heaviside step function with
$\Theta(x) = 1$ if $x>0$ and $0$ otherwise.
Since $\dl$ also diverges with $l$, we cannot simply
neglect the integral over the domain $(0,\dl)$.
If $\dl> z_l$, the above recursion has the solution
\begin{eqnarray}
z_l = (a/\alpha)^{1/\alpha}
\end{eqnarray}
which is of the order of the mean of $f_g$ and does not depend on $l$. This shows that it is not possible
to have a diverging solution that increases more slowly than the handicaps for $b>1$. 
Furthermore, if $z_l$ saturates to a finite number,
$Q_l(x)$ cannot be approximated as a $\delta$-function and the MFA
does not apply. 
Rather, if $z_l \ll \dl$ all events would asymptotically
be records with probability 1 and 
$Q_l(y) \rightarrow f(y)$ as $l\rightarrow \infty$. 
We will return to this scenario in the next subsection.

For now, let us assume that $z_l > \dl$ and ask
when a diverging solution can exist. We 
start from \eref{Eq:dzdl} with $b>1$.
If $\alpha \ge 1$, diverging $z_l$ implies $z_l^{1-\alpha} \ll \dl$,
which leads to the contradictory relation  $z_{l+1} < z_l$.
Thus, for $\alpha  \ge 1$ no diverging solution is possible and asymptotically
all random variables in the background process are records.

The analysis for the case $\alpha <1$ is similar to that
in \sref{Sec:dr_g}, except that now $b > 1$. 
Assuming $z_l^\alpha \approx a (A l)^\gamma$, we arrive at
\eref{Eq:Agammaz}. By the same reasoning as in \sref{Sec:dr_g}, $\gamma$ cannot be larger than 1. If
$\gamma = 1$, we have either $\alpha < 1/b$ with $A=1$ or
$\alpha = 1/b$  with $A$ being the solution of
\eref{Eq:Amplitude} with $b>1$.
In contrast to the case $b<1$, however, \eref{Eq:Amplitude} for $b>1$ does not have 
a positive solution if $c$ is larger than a `threshold'
\begin{eqnarray}
c_t =  a^b \left (1-b^{-1}\right )^{b-1}.
\end{eqnarray}
Thus we expect that $z_l$ diverges as $l^{1/\alpha}$ for $c<c_t$ and saturates to a finite value
for $c>c_t$.
On the other hand, assuming $\gamma<1$ as in \sref{Sec:dr_g}, we get 
$\gamma = (1-b)/(1-1/\alpha)$. However, the condition $\gamma <1$ 
implies $\alpha < 1/b$, and we have seen above that for this case a
solution with $\gamma = 1$ exists. 
Hence the solution with $\gamma <1$ can at best
describe the subleading behavior. 

To summarize, the asymptotic behavior of $z_l$ as predicted by the MFA
for $\alpha \neq 1/b$ is,
\begin{equation}
\frac{z_l^\alpha}{a} = \cases{l - o(l), & $\alpha < 1/b$,\\
\mathrm{finite}, & $\alpha >1/b$,
}
\label{Eq:zl_bl}
\end{equation}
and for $\alpha = 1/b$,
\begin{eqnarray}
\label{Eq:1storder}
\frac{z_l^\alpha}{a} = \cases{ A_b(c)l - o(l), & $c\le c_t$,\\
\mathrm{finite}, & $c>c_t$.
}
\end{eqnarray}

\subsection{\label{Sec:StBi}Stochastic bistability}
The value of the amplitude $A_b(c)$ in \eref{Eq:1storder} is finite at $c=c_t$, which is
suggestive of a first order phase transition as a function of $c$. We
will see in this subsection that such a transition indeed exists, but
its character is importantly modified by fluctuations that have been
neglected in the MFA. 
To make the point clear, we limit ourselves to the distribution
$F_g(x) = 1 - \exp(-x^{1/b})$ with $a=1$
which was anticipated to exhibit a phase transition at $c=c_t$. 
As before, we consider only the asymptotic regime.

\begin{figure}[t]
\centering
\includegraphics[width=0.7\textwidth]{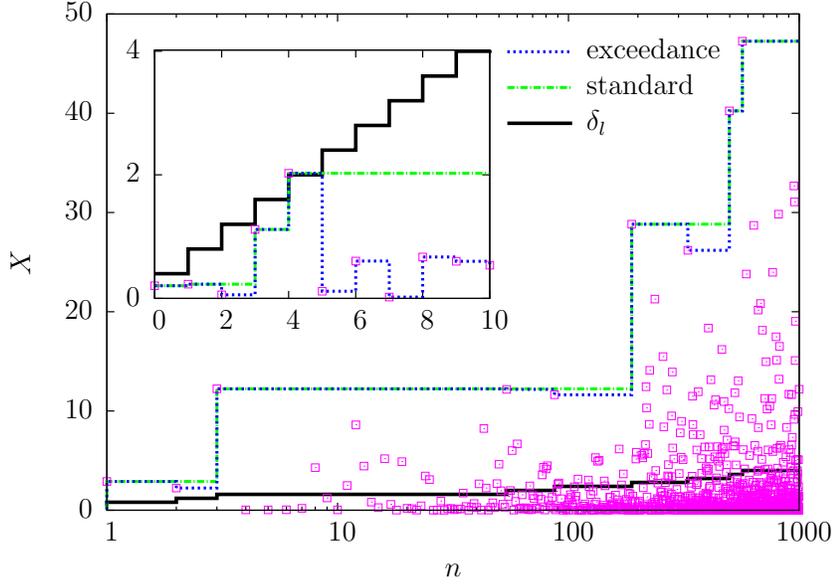}
\caption{\label{Fig:StBi}  A sample path of $\delta$-exceedance
records (blue dotted line) for $b=1/\alpha=2$ and $c =0.4$ on a semi-logarithmic scale. There are 9 new records in the figure.
For comparison, the corresponding standard records (green dot-dashed line) and the handicaps $\dl$ (black 
solid line) are also drawn. 
The random variables of the background process are represented by open squares.
In this case, the $\delta$-exceedance records are very close to the standard records.
Inset: Another sample path for the same parameter sets. In this path, 
the $\delta$-exceedance records are smaller than $\dl$. 
As in the main figure, there are 9 $\delta$-exceedance records which however occur on a much shorter time scale ($n=9$, all events are records).
}
\end{figure}
Suppose that the $l$'th record happens to be smaller than
$\dl$, $Y_l < \delta_l$. As a consequence of the definition (\ref{Eq:Process})
and the fact that the support of the distribution $F(x)$ is limited to
the positive real line,  
the next background event following the $l$'th record is then a record
with probability 1. Since its value $Y_{l+1}$ is an unconstrained draw from the
background distribution, the probability that $Y_{l+1}$ is larger than
$\delta_{l+1}$
is 
\begin{eqnarray}
P_{l+1}^> = 1-F_g(\dlp ) \approx \exp(-c^{1/b} l^{(b-1)/b}),
\end{eqnarray}
which is very small for large $l$. Thus we see that the process is
effectively trapped in the state $Y_k < \delta_k$, where all events
are `records' drawn from the background distribution. As $\delta_k$ increases with
$k$, the corresponding probability $P_k^>$ decreases
further for $k > l+1$, and the expected time until the process for the
first time reverts to $Y_k > \delta_k$ is larger than $1/P_{l+1}^>
\sim \exp(c^{1/b} l^{(b-1)/b}) \gg l$. 
On the other hand, if $Y_l \sim z_l \sim l^b$ for sufficiently
large $l$ as predicted by the MFA in \sref{Sec:ir_mf}, $Y_{l}$ remains
larger than $\dl$ with high probability because $\dl$ only
increases as $l^{b-1}$. 

We conclude that the sample paths of the process segregate into two
subpopulations, a stationary population in which all events of the background process are
records and a diverging population where $Y_l$ grows more rapidly than $\dl$. 
\Fref{Fig:StBi} shows a realization of the segregation phenomenon
due to initial fluctuations. 

To account for this behavior, we approximate the distribution $Q_l(x)$ by a 
sum of two contributions,
\begin{eqnarray}
Q_l(x) \approx \Qs f_g(x) + \left (1 - \Qs \right ) \delta(x-\tilde z_l),
\label{Eq:b2asym}
\end{eqnarray}
where $\tilde z_l$ should diverge faster than $\dl$ and
$\Qs$ is the limiting value of the probability $\Qsl$
that $Y_l$ is smaller than $\dl$,
\begin{eqnarray}
\Qs = \lim_{l\rightarrow \infty} \Qsl \equiv  \lim_{l\rightarrow \infty}
\mathbb{P}[Y_l < \dl].
\label{Eq:QsDef}
\end{eqnarray}
This quantity measures the relative weight of the stationary subpopulation of sample paths and will serve
as an order parameter for the phase transition in the following.

Plugging \eref{Eq:b2asym} into \eref{Eq:dzdla}
and keeping the leading terms, we get
\begin{eqnarray}
\tilde z_{l+1} - \tilde z_{l} 
= \frac{a}{\alpha} \tilde z_{l+1}^{1 - \alpha} - \dl,
\end{eqnarray}
where we approximate $\int_0^\dl (\dl - x) Q_l(x) dx \approx \Qs \dl$.
Hence, the diverging solution found in \sref{Sec:ir_mf} actually
describes the behavior of $\tilde z_l$. Note that the mean $z_l$ and
standard deviation $\sigma_l$ of $Q_l(x)$ are related to $\tilde z_l$
by ($0<\Qs<1$)
\begin{eqnarray}
z_l \approx \tilde z_l\left (1 - \Qs \right ) ,\quad
\sigma_l \approx \tilde z_l\sqrt{\Qs \left ( 1 - \Qs \right )} ,
\end{eqnarray}
which yields the relation
\begin{eqnarray}
\frac{z_l}{\sigma_l} \approx \sqrt{\frac{1}{\Qsl}-1}\equiv  R_l .
\label{Eq:Rzs}
\end{eqnarray}
In this context, the absence of a solution for the prefactor $A$ 
for $c>c_t$ can be interpreted as $Q^{(s)} = 1$. Since the prefactor $A$ 
at $c = c_t$ is $A^\ast = (1-b^{-1})^b a^b$, we expect that $Q^{(s)}$
is strictly smaller than 1 even at $c=c_t$. Hence we can conclude that a discontinuous 
transition in terms of $Q^{(s)}$ occurs at $c=c_t$.

\begin{figure}[t]
\centering
\includegraphics[width=0.7\textwidth]{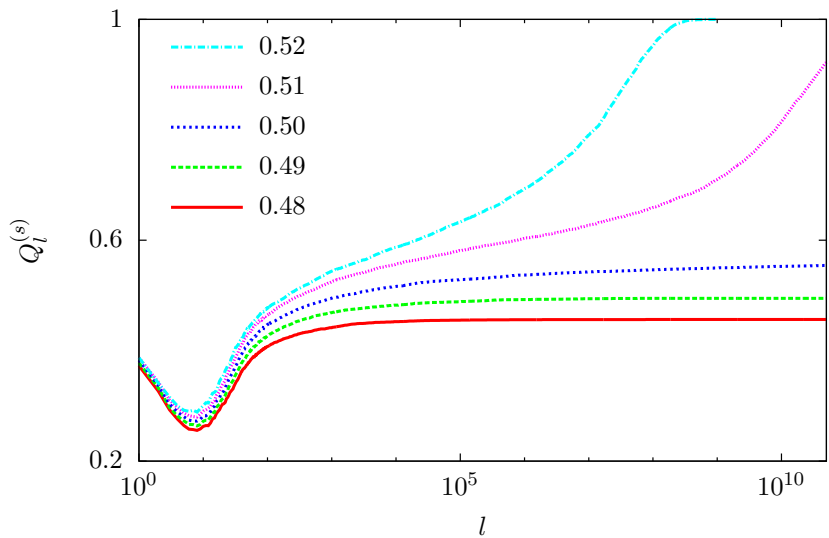}
\caption{\label{Fig:zlnon}  Semi-logarithmic plot of 
$\Qsl$ vs. $l$ for $b=1/\alpha=2$ and
$\dl = c (l+1)$ for $c$ around the threshold value $c_t =\frac{1}{2}$.
}
\end{figure}
To support the above theory, we numerically studied the case with
$b=1/\alpha=2$.
For this case, $c_t=\frac{1}{2}$ and 
\begin{eqnarray}
A = \frac{1+\sqrt{1-2c}}{2},
\label{Eq:A2}
\end{eqnarray}
which is the solution of $A - A^2 = c/2$.
Note that we have only taken the larger solution, expecting
that $A$ is a continuous function of $c$ for $0\le c<c_t$.
We first check if $\Qs$ shows a discontinuity at the 
threshold value $c_t=\frac{1}{2}$. In \fref{Fig:zlnon}, $\Qsl$ is
depicted as a function of $l$ for
$c$ around the threshold value. Each curve is the result of
$10^4$ independent runs. As anticipated, there is a clear
indication of a discontinuous jump at $c=c_t = \frac{1}{2}$.

\begin{figure}[t]
\centering
\includegraphics[width=0.7\textwidth]{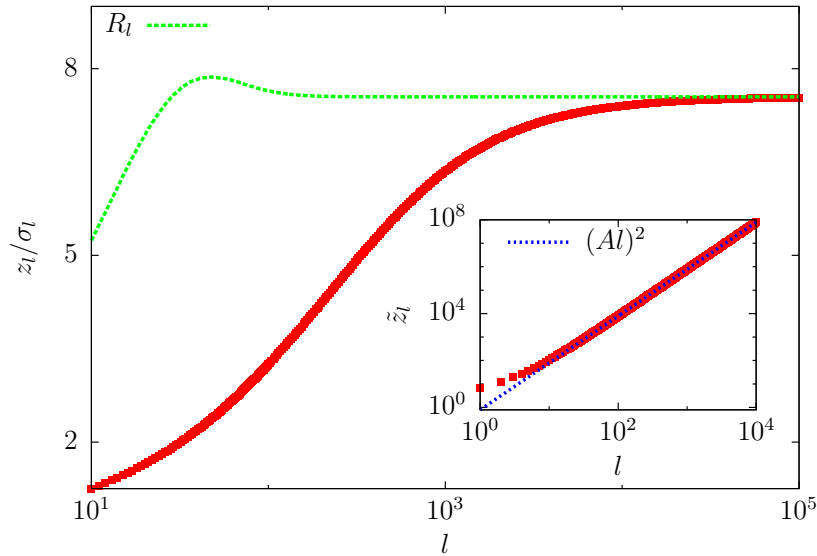}
\caption{\label{Fig:non}  Semi-logarithmic plot of 
$z_l/\sigma_l$ (symbols) vs. $l$ for $b=1/\alpha=2$ and
$\dl = c (l+1)$ with $c=0.2$. 
For comparison, $R_l$ defined in \eref{Eq:Rzs} 
is also drawn (top curve).
Inset: Double logarithmic plot of $\tilde z_l \equiv z_l/(1-\Qsl)$ (symbols) vs. $l$ for $b=1/\alpha=2$ and $c=0.2$.
The straight line is a plot of $(Al)^2$ against $l$ with $A$ 
given by \eref{Eq:A2}.
}
\end{figure}

Next, we  check if \eref{Eq:b2asym} is a valid assumption by
comparing $z_l/\sigma_l$ with $R_l$ in \fref{Fig:non}.
The asymptotic behaviors of both quantities are indeed in good agreement with 
each other.
The inset of \fref{Fig:non} compares $\tilde z_l \equiv z_l/(1-\Qsl)$ with
the anticipated asymptotic behavior $(Al)^2$ with $A$ given by \eref{Eq:A2}.
This also shows an excellent agreement. Although we only present 
data for $c=0.2$, similar agreement is observed in a wide range
of $c$ for $c\le \frac{1}{2}$.

As can be seen from \fref{Fig:non}, $R_l$ saturates in a rather short time.
This indicates that the asymptotic behavior of $\Qsl$ is almost determined by
the fluctuation of $Y_l$ when $l$ is small; see also \fref{Fig:StBi}.
That is, the analysis of the asymptotic
behavior of $Q_l(x)$ and corresponding quantities cannot give much information
about $\Qs$, and we do not think $\Qs$ is universal in the sense that
it is only determined by
the leading behavior of $\dl$.
For example, if $\dl = c (l+l_0)^{b-1}$ with very large $l_0$, $\dl$ for
small $l$ is at least $c l_0^{b-1}$ which means $\Qsl$ is almost 1
for any $\alpha$.
Hence, we have
to resort to numerical analysis to find $\Qs$.

\begin{figure}[t]
\centering
\includegraphics[width=0.7\textwidth]{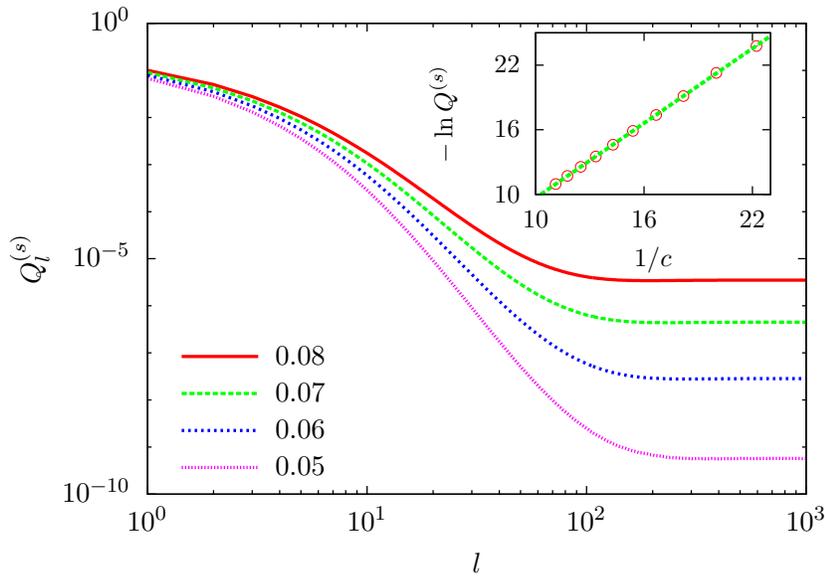}
\caption{\label{Fig:sat}   
Plots of $\Qsl$ vs. $l$ for various values of $c$ on a double-logarithmic scale.
Inset: Plot of $-\ln \Qs$ vs. $1/c$.
The straight line is the result of a linear fit.
The number of independent runs for each data set is $2.5\times 10^{12}$.
}
\end{figure}

Since $\Qs =0$ when $c=0$, it is an interesting question how
$\Qs$ approaches zero as $c \rightarrow 0$.
We investigated the behavior of $\Qs$ for small $c$ via simulations. 
As \fref{Fig:sat} shows, $\Qs$ decreases quite fast for small $c$,
which suggests a form
\begin{eqnarray}
\label{Eq:chifit}
\Qs(c) \sim \chi_1 \exp(-\chi_2/c)
\end{eqnarray}
with two parameters $\chi_1$ and $\chi_2$. 
If this is the case, a plot of $-\ln \Qs$ as a function of $1/c$ should be
well fitted by a straight line. Indeed, as the inset of \fref{Fig:sat} 
shows, a linear function well approximates the data with  
parameter values $\ln \chi_1 \approx 1.9$ and $\chi_2 \approx 1.15$.

\begin{figure}[t]
\centering
\includegraphics[width=0.7\textwidth]{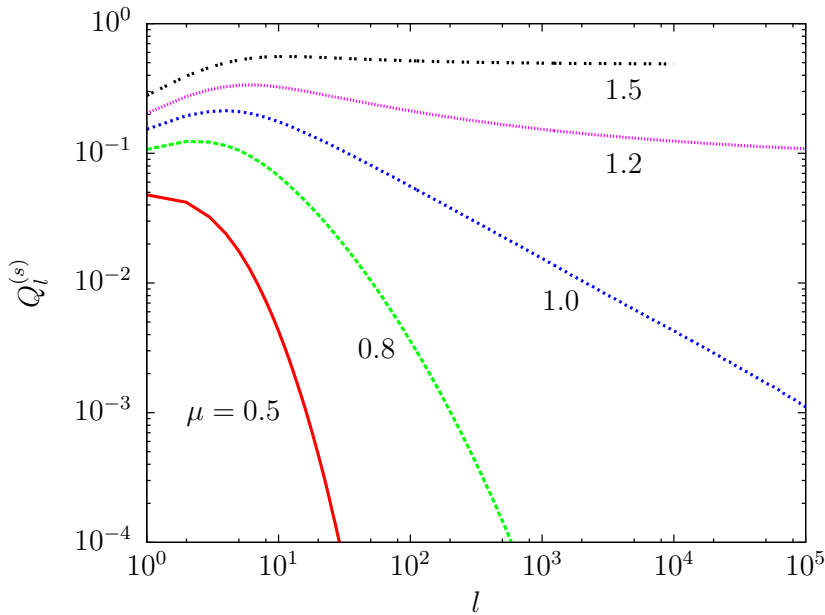}
\caption{\label{Fig:bistab}   
Plots of $\Qsl$ vs. $l$  for $b=2$ and $c=1$ 
on a double-logarithmic scale for Fr\'{e}chet class distributions. 
The values of the power law exponent $\mu$ in this figure are 0.5, 0.8, 1.0, 
1.2, and 1.5 (bottom to top), respectively.
As predicted by the theory,
$\Qsl$ saturates to a finite value when $\mu > 1/(b-1) = 1$.
}
\end{figure}
It should be clear from the above discussion that the stochastic
bistability scenario with a nonzero $Q^{(s)}$ is not restricted to $\alpha = 1/b$ but should
apply also for $\alpha < 1/b$, where the MFA predicts standard record
behavior with $z_l^\alpha \approx al$. In contrast to the case
$\alpha = 1/b$, however, for $\alpha < 1/b$ we expect $Q^{(s)}$  
to be smoothly increasing function of $c$ that approaches unity only
asymptotically for large $c$. We have checked numerically that this is
indeed the case, and found that the behavior of $Q^{(s)}$ for small
$c$ is again well described by the functional form (\ref{Eq:chifit}). 

\subsection{Fr\'echet class}

For $\Qs$ to remain zero for $c > 0$, the 
probability that $Y_l > \delta_l$ under the condition that
$Y_{l-1} < \dlm$ should not be negligibly small for large $l$. 
For the Gumbel (and Weibull) classes, this scenario is clearly not feasible 
as we have seen and only the Fr\'echet class might allow for such a possibility.
Let us consider $F_f(x)$ with $a=1$. The probability of interest is 
$P_l^> \sim l^{-\mu(b-1)}$, hence such an event would happen 
after $m \sim l^{\mu(b-1)}$ records. If $\mu (b-1)<1$,
$l \gg m$ for sufficiently large $l$ and 
the process can escape from the stationary regime before 
$P_l^>$ changes substantially.
Thus, we expect $\Qs=0$ if $\mu < 1/(b-1)$.  

To confirm this argument, we performed simulations for $b=2$ and
various $\mu$. For convenience, we fix $c=1$. 
\Fref{Fig:bistab} depicts $\Qsl$ as a function of $l$ for $\mu = 0.5$, 0.8,
 1.0, 1.2, and 1.5. As anticipated, the long time
behavior of $\Qsl$ changes qualitatively at $\mu_c = 1/(b-1) = 1$.
As a final remark, $\Qsl$ seems to exhibit a power-law decay 
at $\mu = \mu_c$ with a power close to 0.5.

\section{\label{Sec:Walk}Length of adaptive walks}
Using the results of the previous sections, we analyze the length of adaptive 
walks when a population starts from a low fitness genotype $\seq_0$.
As was extensively discussed in Ref.~\cite{PSNK2015}, 
we can assume that every step increases the mutational distance to
$\seq_0$, and that all possible neighboring genotypes of higher fitness from
genotype $\seq_k$ are located in the `forward' direction at distance
$k+1$. In the initial state $\seq_0$ the number of forward neighbors
is $L$, and after $l$ steps it reduces to $L-l$.
Let $Q_l(y,L)$ denote the probability density that the random
part, i.e., $\eta_{\seq_l}$ in \eref{Wk} is $y$ provided that the walker takes at least
$l$ steps. Then the recursion relation for $Q_l(y,L)$ is found to be 
\begin{eqnarray}
\label{Eq:QlL}
Q_{l+1}(y,L) = f(y) \int_{-\infty}^{y+\dl} Q_l(x,L)\frac{1-F(x-\dl)^{L-l}}{1-F(x-\dl)} dx,
\end{eqnarray}
where the term $F(x-\dl)^{L-l}$ accounts for the possibility that
the walker stops because none of the $L-l$ forward neighbors is of
higher fitness \cite{PSNK2015}. This term is absent in the
corresponding recursion relation \eref{Eq:Ql} for the distribution
$Q_l(x)$ of the $\delta$-exceedance
record process, and hence the two distributions are related by 
\begin{eqnarray}
Q_l(x) = \lim_{L\rightarrow \infty} Q_l(x,L).
\end{eqnarray}
 Denoting by $H_l$ and $P_l$ the probability that the walker
takes at least $l$ steps and that the walker stops at the $l$'th step, respectively,
we can write
\begin{eqnarray}
H_l &= \int_{-\infty}^\infty Q_l(x,L) dx,\\
P_l &\equiv H_l-H_{l+1}=\int_{-\infty}^\infty Q_l(x,L) F(x - \dl)^{L-l}dx,
\label{Eq:Pstop}
\end{eqnarray}
which are used to calculate the mean walk distance
\begin{eqnarray}
D_\mathrm{RAW} =\sum_{l=0}^L l P_l.
\label{Eq:Draw}
\end{eqnarray}

\subsection{\label{Sec:deltaQ} Gumbel class}
In this subsection, we will calculate $D_\mathrm{RAW}$ for the Gumbel class
using the results of \sref{Sec:bs} and \sref{Sec:bl}. 
When $b>1$, we have shown that the record process gets trapped in a state
where all entries are records with a non-zero probability $\Qs$. For 
the adaptive walk this means that every randomly chosen neighboring
genotype in the forward direction is of higher fitness, and the walk
therefore attains the maximal possible length $l = L$. On the other
hand, with probability $1- \Qs$ the walk behaves similar to the case
$b < 1$ where, as we will show below, the walk length increases only
logarithmically with $L$. Thus the distribution of walk lengths for 
$b > 1$ is bimodal, with the mean walk length being
dominated by the peak at $l = L$ and hence $D_\mathrm{RAW} \sim O(L)$. 

To calculate $D_\mathrm{RAW}$ with $b<1$, we employ the following approximation
scheme. Since $Q_l(x)$ can be understood as the density of the random
part at the $l$'th step conditioned on the walker taking at least $l$ steps
irrespective of $L$, we approximate $Q_l(x,L)$ as 
$H_{l} Q_l(x)$, which in turn gives 
\begin{eqnarray}
P_l &\approx H_{l} \int_{-\infty}^\infty Q_l(x) F(x-\dl)^{L-l} dx.
\label{Eq:PstopA}
\end{eqnarray}
In particular, when $Q_l(x) \approx \delta(x-z_l)$, $P_l$ becomes
\begin{eqnarray}
P_l =H_l - H_{l+1} = H_{l} F(z_l-\dl)^{L-l}.
\end{eqnarray}
We have shown in \sref{Sec:bs} that  
$F(z_l - \dl)$ can generally be approximated as $\exp[-e^{-(Al)^{\gamma}}]$,
with certain numbers $A>0$ and $\gamma \le 1$.
Since $H_l$ is not expected to be
significantly different from $H_{l+1}$, we can treat $H_l$ as a
differentiable function of $l$. Hence, for sufficiently large $l$ and
$L \gg l$ we get
\begin{eqnarray}
\frac{dH(l)}{dl} = - H(l) \exp\left [-Le^{-(Al)^{\gamma}} \right ]
\end{eqnarray}
with the solution
\begin{eqnarray}
H(l) \approx \exp\left [ -\int_0^l \tP(x)dx \right ],
\end{eqnarray}
where 
\begin{eqnarray}
\tP(x) = \exp\left [-Le^{-(Ax)^{\gamma}} \right ].
\end{eqnarray}
Since $P_l \approx - \frac{dH}{dl}$, we arrive at
\begin{eqnarray}
D_\mathrm{RAW} \approx \int_0^\infty y \tP(y) \exp \left [ -\int_0^y \tP(x) dx
\right ] dy,
\label{Eq:Dintori}
\end{eqnarray}
where we have assumed that $P_l$ is negligible if $l = O(L)$.

\begin{figure}[t]
\centering
\includegraphics[width=\textwidth]{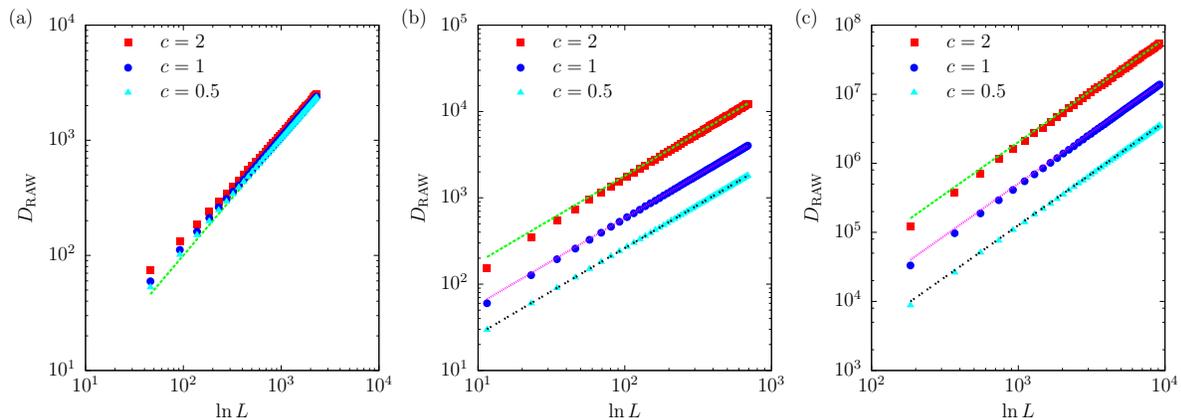}
\caption{\label{Fig:Sb}  Double-logarithmic plots of the adaptive walk length $D_\mathrm{RAW}$ vs. $\ln L$ 
for $b=\frac{1}{2}$ and Gumbel-class random variables with (a) $\alpha=1$, (b) $\alpha=2$, and (c) $\alpha=4$. 
Straight lines are the predicted leading term in \eref{Eq:Dint}.
}
\end{figure}

By the change of variables $y =w+ ( \ln L)^{1/\gamma}/A$ and $x = z+(\ln L)^{1/\gamma}/A$, we get
\begin{eqnarray}
\fl
D_\mathrm{RAW} \approx
\int_{-\infty}^\infty dw \left [ \frac{ ( \ln L)^{1/\gamma}}{A} + w\right ] 
\exp\left [ - e^{-w/K} - \int_{-\infty}^w \exp[-e^{-z/K}] dz \right ],
\end{eqnarray}
where $K = (\ln L)^{1/\gamma-1}/(A\gamma)$. After further substitutions
$e^{-w/K} = x$ and $e^{-z/K} = t$, we obtain
\begin{eqnarray}
D_\mathrm{RAW} \approx
\int_{0}^\infty dx \left [ \frac{ ( \ln L)^{1/\gamma}}{A} -K \ln x \right ] 
\frac{d}{dx}
\exp \left [- KE_1(x) \right ],
\end{eqnarray}
where 
\begin{eqnarray}
E_1(x) = \int_x^\infty \frac{e^{-t}}{t} dt
\end{eqnarray}
is the exponential integral function.
Thus, we have
\begin{eqnarray}
D_\mathrm{RAW} &\approx 
 \frac{ (\ln L)^{1/\gamma} }{A}+  \frac{1}{A\gamma} (\ln L)^{1/\gamma -1} \K(K),
\label{Eq:Dint}
\end{eqnarray}
where
\begin{eqnarray}
\K(K) \equiv -\int_0^\infty \ln x \frac{d}{dx} \exp\left [-KE_1(x)\right ] dx.
\end{eqnarray}
For $A=\gamma=1$ the numerical value of $\K(K)$ is 
$1.099~124$ which coincides with the exact result for $c=0$~\cite{FL1992}.
However, since our analysis neglects the effect of fluctuations, the subleading behavior
of \eref{Eq:Dint} cannot generally be expected to be exact. 
The leading order behavior
$D_\mathrm{RAW} \approx (\ln L)^{1/\gamma}/ A$
with $A$ and $\gamma$ given in \eref{Eq:A} and \eref{Eq:gamma},
respectively, is compared to simulations in \fref{Fig:Sb}, showing
excellent agreement.

\subsection{Weibull class}

Now we consider the bounded distribution of Weibull type $F_w$.
Since the case $b>1$ is trivial, we limit ourselves to the case $b<1$.
For a rough estimate of the adaptive walk length, we ask at which
value of $l$ the stopping probability $F(x-\dl)^{L-l}$ in \eref{Eq:QlL} has an appreciable
magnitude when evaluated near the mean record value $z_l$ \cite{PSNK2015}. We have seen in
\sref{Sec:dr_w} that the approach of $z_l$ towards the upper
boundary $r$ is determined by the behavior of the handicaps, in the
sense that $r-z_l \sim \dl$. Thus using \eref{Eq:boundzlsl} we have
\begin{eqnarray}
\label{Eq:Fwest}
F_w(z_l - \dl)^{L-l} \approx \exp\left [-\left(\frac{(1+\nu)c}{\nu
      r}\right)^{1/\nu} \frac{L-l}{l^{(1-b)/\nu}}\right ].
\end{eqnarray}
Let us first assume that $l \sim L^\xi$ for some $\xi < 1$. If $1-b
> \nu$ the right hand side of \eref{Eq:Fwest} becomes of order unity for $L
\to \infty$ if $\xi = \nu/(1-b)$, leading to the prediction that
\begin{eqnarray}
\label{Eq:DRAW-Weibull}
D_\mathrm{RAW} \sim L^{\nu/(1-b)}.
\end{eqnarray} On the other hand, if $1-b <
\nu$ the right hand side will vanish with $L \to \infty$ for any $\xi
< 1$, which implies that the walk length must be $O(L)$ to leading
order. Indeed, taking $L-l = L^{\xi'}$ we see that \eref{Eq:Fwest}
approaches a nonzero limit if $\xi' = (1-b)/\nu$, and we
conclude that $D_\mathrm{RAW} = L - O(L^{(1-b)/\nu})$ for any $c>0$
once $\nu > 1-b$.

For the case of $1-b > \nu$ 
our simulations show a rather systematic, though small, deviation
from the prediction \eref{Eq:DRAW-Weibull}, which indicates
the need for a more careful analysis. To get a more accurate expression of $D_\mathrm{RAW}$ for
$\nu< 1-b$, we start from the approximation \eref{Eq:PstopA}, which now takes
the form
\begin{eqnarray}
\frac{P_l}{H_l} &\approx \int_0^r dx F_w(r-x-\dl)^{L} Q_l(r-x)
\approx \cP\left (\frac{K}{l}\right ),
\end{eqnarray}
where $ K = (cL^{\nu}/\bdw)^{1/(1-b)} $, 
\begin{eqnarray}
\cP(x) = \int_0^\infty dy \exp\left [-x^{(1-b)/\nu}(1+y)^{1/\nu} \right ] g(y) dy
\end{eqnarray}
and the function $g(y)$ was defined in \eref{Eq:gy}.
Since we expect $D_\mathrm{RAW} \sim o(L)$, we dropped the $l$ in $F^{L-l}$.

Although we do not know an explicit form for $g(y)$, we can still obtain
the asymptotic behavior of $\cP$ for small and large $x$.
When $x \ll 1$, we can approximate $\cP(x)$ as
\begin{eqnarray}
\cP(x) \approx 1 -  x^{(1-b)/\nu} \int_0^\infty \left
(1+y \right )^{1/\nu} g(y) dy.
\label{Eq:cPapp}
\end{eqnarray}
When $x\gg 1$, the integral will be dominated by small $y$, so 
we can approximate
\begin{eqnarray}
\cP(x) &\approx \exp\left [-x^{(1-b)/\nu} \right ] \int_0^\infty dy \exp\left [ - \frac{x^{(1-b)/\nu}}{\nu }y\right ] g(y) \nonumber \\
&\sim \exp \left [ -x^{(1-b)/\nu}\right ]x^{-(1-b)/\nu^2} \nu^{1/\nu} ,
\label{Eq:cPsmall}
\end{eqnarray}
where we have used that $\G_l(z/\dl) \sim z^{-1/\nu}$ for large $z$ (see
\sref{Sec:dr_w}). 

Following the same line of reasoning as in \sref{Sec:deltaQ},
we obtain the approximate formula
\begin{eqnarray}
\label{Eq:cPlarge}
D_\mathrm{RAW} &\approx \int_0^\infty dy y \cP\left (\frac{K}{y}
\right ) 
\exp\left [-\int_0^y \cP\left ( \frac{K}{x}\right ) dx \right ]\nonumber\\
&= K \int_0^\infty \frac{dz}{z}
\exp\left [ - K \int_z^\infty \cF(t) dt 
+ \ln \left \{ K\cF(z) \right \}\right ],
\end{eqnarray}
for the walk length, where $\cF(x) = \cP(x)/x^2$.
To estimate the integral, we look for the saddle point $z_c$ that maximizes
the argument of the exponential function, which satisfies the equation
\begin{equation}
0 = \left . \frac{d}{dz} \left [ K \int_z^\infty \cF(t) dt - \ln \cF(z) \right ]\right |_{z=z_c}
=  \left . -K\cF(z_c)  - \frac{d\ln \cF(z) }{dz} \right |_{z=z_c}.
\label{Eq:zc_eq}
\end{equation}
Since $K$ is large, $z_c$ should be either very small or very large.
Let us assume that $z_c$ is very small. Approximation of $\cP$ for small
$x$ in \eref{Eq:cPapp} gives $\cF(z) \sim z^{-2}$, which
suggests $-K/z_c^2 + 2/z_c = 0$ or $z_c \sim K$. Thus, assuming
that $z_c$ is very small leads to a contradiction. 

Let us now investigate if a large $z_c$ solution exists.
When $z$ is large, $\cF(z) \approx C_0 z^{-v}\exp(-z^\epsilon)$ with $\epsilon = (1-b)/\nu$,
$v = 2 + (1-b)/\nu^2$, and $C_0 = \nu^{1/\nu}$.
Thus, we can approximate $\cF'(z_c)/\cF(z_c) \approx -\epsilon z_c^{\epsilon-1}$ which, together with \eref{Eq:zc_eq}, gives
\begin{equation}
K \cF(z_c) \approx K C_0 z_c^{-v} \exp(-z_c^\epsilon)
\approx \epsilon z_c^{\epsilon-1}.
\label{Eq:zc_sol}
\end{equation}
Thus, we find the leading behavior of $z_c$ as
\begin{equation}
z_c \sim \left ( \ln K \right )^{\nu / (1-b)},
\label{Eq:zck}
\end{equation}
which is consistent with the large $z_c$ assumption.

Using \eref{Eq:zck}, we can now find $D_\mathrm{RAW}$. Since
\begin{equation}
\exp\left [ - K \int_z^\infty \cF(t) dt 
+ \ln \left \{ K\cF(z) \right \}\right ]
= \frac{d}{dz} \exp\left [ - K \int_z^\infty \cF(t) dt \right ],
\end{equation}
and the exponential function is dominated by the region around $z=z_c$,
we approximate
\begin{eqnarray}
D_\mathrm{RAW} &\approx \frac{K}{z_c} \int_0^\infty dz
\frac{d}{dz} \exp\left [ - K \int_z^\infty \cF(t) dt \right ]\nonumber \\
&= \frac{K}{z_c}
\sim \frac{K}{(\ln K)^{\nu/(1-b)}} = 
\left (\frac{1-b}{\nu} \frac{L}{\ln L}\right )^{\nu/(1-b)},
\label{Eq:WDR}
\end{eqnarray}
where we have used $\int_0^\infty \cF(t) dt = \infty$ (recall
that $\cF(t) \sim t^{-2}$ for small $t$).

\begin{figure}[t]
\centering
\includegraphics[width=0.7\textwidth]{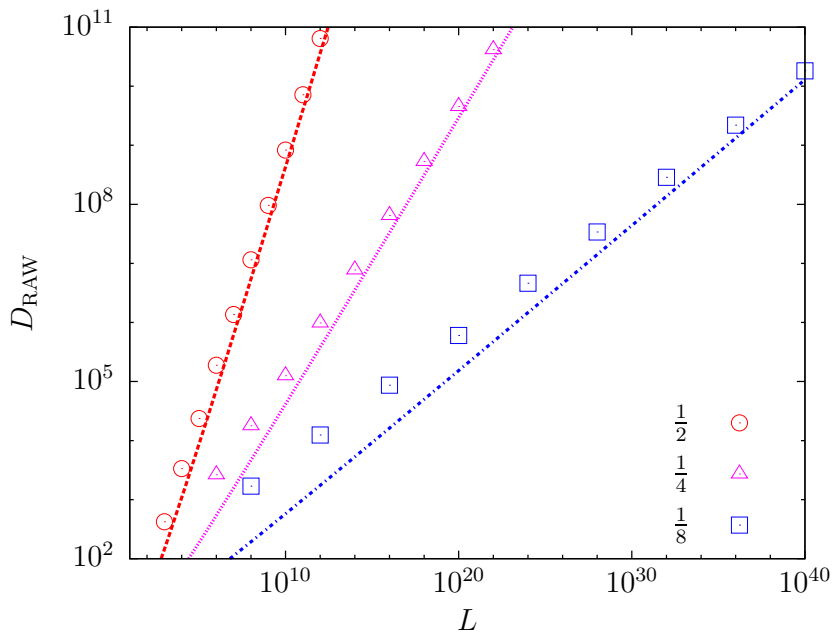}
\caption{\label{Fig:WD}  Double-logarithmic plots of the adaptive walk length $D_\mathrm{RAW}$ vs.
$L$ for $b=\frac{1}{2}$ and distributions belonging to the Weibull class
with $\nu =\frac{1}{2}$ (circles), $\frac{1}{4}$ (triangles),
and $\frac{1}{8}$ (squares). For comparison, the prediction \eref{Eq:WDR} is shown as straight lines.
}
\end{figure}

To confirm this asymptotic behavior, we performed numerical simulations
for $b=\frac{1}{2}$. \Fref{Fig:WD} shows that the asymptotic behavior
for large $L$ is well described by \eref{Eq:WDR}.

\section{\label{Sec:Summary}Summary and conclusions}
In this paper we have investigated a modified record process defined on sequences of i.i.d. 
random variables in which the occurrence threshold for record events is reduced by 
a handicap $\delta_k$ that is a function of the record number $k$. This modification obviously
increases the rate of record occurrence and decreases the magnitude of record values. However, 
similar to other cases where the standard record process is modified by rounding effects or trends
\cite{Franke2010,Wergen2012}, the degree to which the record statistics are altered by the 
handicaps depends strongly on the tail properties of the background process. By allowing the 
handicaps to depend on the record number, we are able to tune their size to match
the increments between the standard record values corresponding to a given background distribution. 
Building on the results obtained in \cite{PSNK2015}, we have thus
uncovered a rich variety of phase-transition like phenomena that emerge from the interplay of the
stochastic record process with the deterministic handicap function $\delta_k$.  

\begin{figure}[t]
\centering
\includegraphics[width=0.7\textwidth]{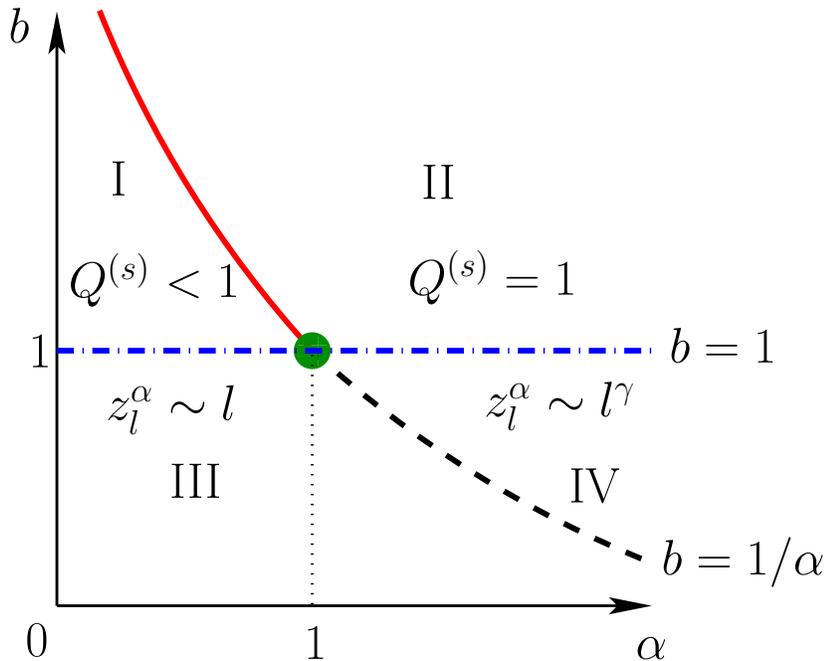}
\caption{\label{Fig:Phase}  Phase diagram summarizing our results for
Gumbel class distributions. Four different regions 
I, II, III, and IV are identified, depending on the behavior of $\tilde z_l$ and
$\Qs$ in \eref{Eq:sum_zlQ}.
}
\end{figure}

When the handicaps increase or decrease according to the power law (\ref{deltak}) with exponent $b-1$,
the distributions whose record  increments  match this behavior are the representatives $F_g$ of the Gumbel
class with exponent $\alpha = 1/b$. In \fref{Fig:Phase} we summarize our findings for this class of distributions 
in the form of a phase diagram in the $(\alpha, b)$-plane. There are four distinct regions separated by the lines
$b=1/\alpha$ and $b=1$. For $b > 1$ (increasing handicap) we have seen that the sample paths of the record process
display a kind of stochastic bistability, which leads to the decomposition of the distribution of record values $Q_l(x)$
into the general form
\begin{equation}
Q_l(x) = \Qs \rho(x) + \left [ 1 - \Qs \right ] \tilde{\rho}_l(x - \tilde z_l ).
\label{Eq:sum_zlQ}
\end{equation}
Here $\rho(x)$ is a probability density with finite mean,
$\tilde z_l$ diverges with $l$ and $\tilde{\rho}_l$ is a distribution with zero mean and a standard deviation
that grows more slowly than $\tilde z_l$, such that the diverging part of (\ref{Eq:sum_zlQ}) becomes concentrated around
$\tilde z_l$ for large $l$. The emergence of a stationary component with weight $\Qs$ 
in a record process that by nature is non-stationary is perhaps
the most remarkable feature of our work. It is well known that records from i.i.d. sequences with an added linear trend become asymptotically
stationary \cite{Ballerini1987}, but the scenario of a first-order-like phase transition in the population of sample paths that we have described
in section \ref{Sec:bl} does not appear to have any counterpart in previous studies of record processes.   

In region I of the phase diagram ($b>1$ and $\alpha<1/b$), $\tilde z_l^\alpha = l - o(l)$ and
$\Qs$ is an increasing function of $c$ that satisfies 
$0<\Qs<1$ for all $c>0$. For small $c$ we have found that $\Qs$ is well approximated
by the function (\ref{Eq:chifit}) which displays an essential singularity  at $c=0$. 
Region II ($b>1$ and $\alpha>1/b$) is characterized by
$\Qs=1$, which means that almost all i.i.d. random variables become records 
when the number of record events is large.
In both regions I and II, $\rho(x)$ in \eref{Eq:sum_zlQ} is asymptotically equal to the density of the 
background process. 

On the borderline between regions I and II which is depicted by the red solid 
curve in \fref{Fig:Phase}, $\Qs$ found to exhibit a discontinuous transition 
as $c$ increases. That is, there is a number $c_t>0$ such that
$\Qs$ is strictly smaller than 1 if $c\le c_t$ while
$\Qs = 1$ if $c > c_t$. When $c \le c_t$, $\tilde z_l^\alpha \approx A_b(c) l$,
where $A_b(c)$ is the (larger) positive solution of \eref{Eq:Amplitude} with
$b>1$. Again, along this curve $\rho(x)$ is the density of the background process
when $\Qs > 0$.

The behavior along the line $b=1$ (blue dot-dashed line in \fref{Fig:Phase})
was the topic of \cite{PSNK2015}, but for completeness we include a discussion of this case in terms of $\Qs$.
For $\alpha<1$, $\Qs = 0$ and $z_l^\alpha = \tilde z_l^\alpha = l - o(l)$ , whereas
$\Qs = 1$ for $\alpha > 1$. 
At $\alpha = b=1$ (represented as a green filled circle in \fref{Fig:Phase}), $\Qs$ changes its behavior discontinuously
from 0 ($c\le a$) to 1 ($c>a$) similar to the behavior along the line $b =1/\alpha$
with $b>1$.  Unlike the case of $b>1$, however, $\rho(x)$ is not equal to the density of the background process
when $\Qs = 1$. Moreover, at the critical point $c=a$, the mean record value increases anomalously slowly 
as $z_l \sim \sqrt{l}$. For further details we refer the reader to \cite{PSNK2015}.

In both regions III and IV ($b<1$), $\Qs = 0$ and $z_l = \tilde z_l$ diverges.
The difference between region III and IV is characterized by
the asymptotic behavior of $z_l^\alpha$.
In region III $z_l^\alpha  = l- o(l)$ like in the standard record process, while
in region IV $z_l^\alpha \sim l^\gamma$ with $\gamma = (1-b)/(1-1/\alpha) < 1$.
On the borderline between regions III and IV, which is represented
by a black dashed line in \fref{Fig:Phase}, $z_l^\alpha = A l
- o(l)$ with a $c$-dependent constant $A$ which is the positive
solution of \eref{Eq:Amplitude} and decreases continuously from $A=1$ at $c = 0$ to 
$A = 0$ for $c \to \infty$. The point $\alpha = b = 1$ is again special in that here the amplitude $A$ vanishes at a finite
value $c = a$ \cite{PSNK2015}. When interpreted in terms of the amplitude $A$ of the diverging component in \eref{Eq:sum_zlQ}, the
transition to the stationary phase is continuous at $\alpha=b=1$ but discontinuous for $1/\alpha = b > 1$. 
 
The behavior for distributions in the Fr\'echet and Weibull classes corresponds roughly 
to that of Gumbel-class distributions with very heavy ($\alpha \to 0$) and light ($\alpha \to \infty$) tails, 
respectively. Specifically, for the Fr\'echet class distributions
the handicap is irrelevant when $b<1$. For $b > 1$ the stochastic bistability scenario 
observed for the Gumbel class applies and $\Qs$ is generally nonzero.  
An exception are heavy tailed distributions with tail exponent $\mu < 1/(b-1)$,
where the fluctuations of the background process are sufficiently strong to overcome the 
increasing handicap and $\Qs = 0$.  This is reminiscent of the problem of records from i.i.d. sequences with a linear trend,
where the record process becomes asymptotically stationary only if the underlying distribution has a finite first moment
\cite{Ballerini1987}.   
For distributions with bounded support belonging to the Weibull class, all but a 
finite number of random variables become records for $b>1$. When $b<1$, the 
approach of the mean record value to the boundary $\bdw$ of the support is dominated by the handicap
in the sense that $\bdw - z_l \sim \dl$, and
the tail behavior of the density of record values is found to be the same
as that of the density $f_w$ of the background process.

An important motivation for our study comes from the connection to adaptive walks in rugged fitness landscapes
with deterministic epistasis in the sense of \cite{Wiehe1997}.
In \sref{Sec:Walk}  the results obtained for the $\delta$-exceedance records were used to quantify
the increase in the mean walk length that is caused by the increasing deterministic fitness profile $g_k$ in 
\eref{Wk}. For $b > 1$ the stochastic bistability of the record process implies that the distribution of walk lengths 
becomes bimodal. With a finite probability $\Qs$, walks traverse the entire fitness landscape and reach the 
maximal possible length $O(L)$. In the case $b < 1$ which corresponds to the biologically important scenario
of diminishing returns epistasis, the effect on the walk length is more subtle and depends sensitively on 
the distribution of the random fitness component. In the Fr\'{e}chet class and in the Gumbel class with $\alpha < 1/b$ the asymptotic behavior 
is $D_\mathrm{RAW} \sim \ln L$ as on an uncorrelated landscape, whereas for light-tailed Gumbel class distributions with 
$\alpha > 1/b$ there is a slight increase in the walk length which now grows as $(\ln L)^{1/\gamma}$ with $\gamma < 1$. By contrast, 
the walk length for the Weibull class grows at least as a power law in $L$ and is given by 
\begin{equation}
D_\mathrm{RAW} \sim
\cases{L& if $\nu > 1-b$ \\ (L/\ln L)^{\nu/(1-b)} & if $\nu \le 1-b$.}
\end{equation}
We note that these results are potentially relevant for the interpretation of 
microbial evolution experiments, where examples of fitness distributions belonging 
to each of the three EVT classes have been identified empirically \cite{Kassen2006,Rokyta2008,Schenk2012,Bank2014}. 

The analysis of adaptive walk length also provides some insight into the statistics of record occurrence times in the $\delta$-exceedance record process, which
we have not explicitly addressed in this work. As was explained in \sref{Sec:AW}, the walk length $D_\mathrm{RAW}(L)$ is expected to be of the same order as the number
of record events up to time $n=L$, a relation that can be made precise for a particular variant of the adaptive walk problem called `simple' adaptive walk in \cite{O2002}. 
This relation reproduces the fact that a finite fraction of random variables are records when $D_\mathrm{RAW}(L) \sim L$, and can be used to estimate the rate of 
record occurrence in the other cases analyzed in this paper. A detailed analysis of the temporal statistics of the $\delta$-exceedance process, including in particular
the question of correlations between record events \cite{Wergen2011,Franke2012}, appears to be an interesting problem for future study.

\ack
We thank Satya Majumdar for useful discussions. 
S-CP acknowledges the support by the Basic Science Research Program through the
National Research Foundation of Korea~(NRF) funded by the Ministry of
Science, ICT and Future Planning~(Grant No. 2014R1A1A2058694);
and by The Catholic University of Korea, Research Fund, 2016.
JK acknowledges the support of DFG within SPP 1590, and the kind hospitality
of the University of Florence during the completion of this work. 
\appendix
\section{\label{App:Der}Derivation of \eref{Eq:exactzl}}
In this appendix, we derive the recursion relations for the mean  $z_l$ and 
the variance $V_l$ of $Q_l(x)$ from \eref{Eq:Ql}. We consider a bounded density $f(x)$ with the 
support $0<x<\bd$, and the case with unbounded support will be obtained
by taking the limit $\bd \rightarrow \infty$.

We first introduce the moment generating function
\begin{eqnarray}
G_l(\lambda) = \int_0^\bd e^{-\lambda x} Q_l(x)dx ,
\label{Eq:Gdef}
\end{eqnarray}
which gives
\begin{eqnarray}
z_l = - \left . \frac{dG_l}{d \lambda} \right |_{\lambda = 0},\qquad
V_l = \left . \frac{d^2 \ln G_l}{d\lambda^2} \right |_{\lambda = 0}.
\end{eqnarray}
For convenience, we introduce
\begin{eqnarray}
F_c(x) = 1 - F(x),\qquad
\tilde F(x) = e^{-\lambda x} F_c(x),
\end{eqnarray}
and we define $F(x) = 0$ if $x<0$.
Later, we will also use the following identity,
\begin{eqnarray}
e^{-\lambda x} f(x) = - \frac{d \tilde F(x)}{dx} -\lambda \tilde F(x).
\end{eqnarray}
Note that $Q_{l+1}(y)$ for $y > \bd - \dl$ is given by
\begin{eqnarray}
\label{Eq:AQl1}
Q_{l+1}(y) = I_0 f(y),
\end{eqnarray}
where 
\begin{eqnarray}
I_0 \equiv \int_0^\bd \frac{Q_l(x)}{1-F(x-\dl)} dx,
\end{eqnarray}
which does not depend on $y$.

We now insert \eref{Eq:Gdef} into the recursion \eref{Eq:Ql}. 
After separating the integration domain followed
by integration by parts, $G_{l+1}(\lambda)$ can be written as
\begin{eqnarray}
G_{l+1}(\lambda) &= \int_0^{\bd-\dl} e^{-\lambda y} Q_{l+1}(y) dy + \int_{\bd-\dl}^{\bd} e^{-\lambda y} Q_{l+1}(y) dy\nonumber\\
&=\int_0^{\bd-\dl} dy e^{-\lambda y} f(y) \int_0^{y+\dl} \frac{Q_l(x)}{F_c(x-\dl)}
+I_0 \int_{\bd-\dl}^\bd dy e^{-\lambda y} f(y) 
\nonumber \\
&= I_1 + I_0 \tilde F(\bd-\dl) 
-\lambda I_0 \int_{\bd-\dl}^\bd dy \tilde F(y),
\label{Eq:Glp1}
\end{eqnarray}
with
\begin{eqnarray}
\fl
I_1 &\equiv  \int_0^{\bd-\dl} dy e^{-\lambda y} f(y) \int_0^{y+\dl} \frac{Q_l(x)}{F_c(x-\dl)} dx \nonumber\\
\fl
&= \int_0^{\bd-\dl} e^{-\lambda y} f(y) dy \int_0^{\dl} Q_l(x) dx
+ \int_0^{\bd-\dl} dy e^{-\lambda y} f(y) \int_{\dl}^{y+\dl} \frac{Q_l(x)}{F_c(x-\dl)} dx\nonumber \\
\fl
&=
\left [ 1-\tilde F(\bd-\dl)  -\lambda \int_0^{\bd-\dl} dy \tilde F(y) \right ]
\int_0^{\dl} Q_l(x) dx
\nonumber \\
\fl
&\quad+ \int_0^{\bd-\dl} dy e^{-\lambda y} f(y) \int_{0}^{y} \frac{Q_l(x+\dl)}{F_c(x)} dx,
\label{Eq:I1}
\end{eqnarray}
where we have changed the variable $x \mapsto x+\dl$.
Changing the order of integration and then integrating by parts, 
the last integral in \eref{Eq:I1}, to be denoted by $I_2$, becomes
\begin{eqnarray}
I_{2}&=\int_{0}^{\bd-\dl} dx \frac{Q_l(x+\dl)}{F_c(x)}
\int_{x}^{\bd-\dl} dy e^{-\lambda y} f(y)\nonumber\\
&= \int_{\dl}^{\bd} e^{-\lambda(x-\dl)} Q_l(x) dx 
- \tilde F(\bd-\dl)
\int_{\dl}^{\bd} dx \frac{Q_l(x)}{F_c(x-\dl)}\nonumber \\
&\quad-\lambda \int_0^{\bd-\dl} dy \tilde F(y) \int_{\dl}^{y+\dl} \frac{Q_l(x)}{F_c(x-\dl)} dx.
\label{Eq:I2}
\end{eqnarray}
The last integral in \eref{Eq:I2}, to be denoted by $I_3$, can be written as
\begin{eqnarray}
I_3 = 
-\lambda \int_0^{\bd-\dl} e^{-\lambda y}\frac{Q_{l+1}(y)}{h(y)}dy  
+\lambda \int_0^{\bd-\dl} dy \tilde F(y)\int_{0}^{\dl} Q_l(x)dx,
\end{eqnarray}
where $h(x) = f(x)/F_c(x)$ is the hazard function.
Hence we get
\begin{eqnarray}
\fl
G_{l+1}(\lambda)  &= \int_{\dl}^{\bd} e^{-\lambda(x-\dl)} Q_l(x) dx  + \int_0^\dl Q_l(x)dx
-\lambda \int_0^{\bd} e^{-\lambda x}\frac{Q_{l+1}(x)}{h(x)}dx \nonumber \\
\fl
&= e^{\lambda \dl} G_l(\lambda) -\lambda \int_0^\bd e^{-\lambda x}\frac{Q_{l+1}(x)}{h(x)} dx 
+ \int_0^{\dl} \left ( 1 - e^{-\lambda(x-\dl)}\right ) Q_l(x)dx ,
\label{Eq:Gexact}
\end{eqnarray}
where we have used 
\eref{Eq:AQl1} for $y>\bd-\dl$.
Note that even if $\bd = \infty$ the above relation is still valid. 

For the first moment, if it exists, we get
\begin{eqnarray}
z_{l+1} &= \left .-\frac{dG_{l+1}}{dy} \right |_{\lambda = 0}
= z_l - \dl + \int_0^\bd \frac{Q_{l+1}(x)}{h(x)} dx
- \int_0^\dl (x-\dl) Q_l(x) dx\nonumber\\
&=\int_0^\bd \frac{Q_{l+1}(x)}{h(x)} dx
+ \int_\dl^\bd (x-\dl) Q_l(x) dx
\label{Eq:zexact}
\end{eqnarray}
and for the second moment $\xi_l$, if it exists, we get
\begin{eqnarray}
\xi_{l+1} &= \left .\frac{d^2G_{l+1}}{dy^2} \right |_{\lambda = 0}
\nonumber
\\&= \xi_l - 2 \dl z_l + \dl^2 
+ 2 \int_0^\bd x \frac{Q_{l+1}(x)}{h(x)} dx
-\int_0^\dl (x-\dl)^2 Q_l(x) dx,
\end{eqnarray}
which gives the recursion relation for the second cumulant, or
variance, $V_l= \xi_l - z_l^2$ as
\begin{eqnarray}
\fl
V_{l+1} - V_l = (z_l-\dl)^2 - z_{l+1}^2 + 2 \int_0^\bd x \frac{Q_{l+1}(x)}{h(x)} dx
- \int_0^\dl (x-\dl)^2 Q_l(x) dx.
\label{Eq:Vl}
\end{eqnarray}

If the support is unbounded on both sides, we get
\begin{eqnarray*}
G_{l+1}(\lambda) &= \int_{-\infty}^\infty e^{-\lambda y} Q_{l+1}(y) dy
= \int_{-\infty}^\infty dx \frac{Q_l(x)}{ F_c(x-\dl)} \int_{x-\dl}^\infty
e^{-\lambda y} f(y) dy \\
&= e^{\lambda\dl }G_l(\lambda)
-\lambda \int_{-\infty}^\infty dx \frac{Q_l(x)}{F_c(x-\dl)} \int_{x-\dl}^\infty
\tilde F(y) dy\\
&=e^{\lambda \dl}G_l(\lambda) -\lambda\int_{-\infty}^\infty e^{-\lambda x} \frac{Q_{l+1}(x)}{h(x)} dx,
\end{eqnarray*}
which is \eref{Eq:Gexact} without the last integral.

\section{\label{App:Variance}Analysis of the variance for decreasing
handicaps in the Gumbel class}
In this appendix, we will show that the MFA used in \sref{Sec:bs} 
is valid in the sense that $V_l/z_l^2 \rightarrow 0$ as $l \rightarrow \infty$.
Since the variance $V_l$ is not neglected, we relax the MFA 
in such a manner
that $Q_l(x) = q[(x-z_l)/\sigma_l]/\sigma_l$, where $\sigma_l \equiv \sqrt{V_l}$ is
the standard deviation and $q(y)$ is supposed to be independent of $l$.
Note that the mean and variance of $q(y)$ are 0 and 1, respectively.
We are still assuming that $Q_l(x)$ has a well-defined 
steady state distribution when $x$ is appropriately rescaled.
Recalling that the hazard function for the Gumbel class distributions is $h(x) = \alpha x^{\alpha -1}/a$, see \eref{Eq:hazard}, 
we may thus approximate the integral terms on the right hand sides of \eref{Eq:zexact} and \eref{Eq:Vl} as 
\begin{eqnarray}
\int dx x^{n-\alpha} Q_{l+1}(x)
&= \int dt (z_{l+1} + \sigma_{l+1} t)^{n-\alpha} q(t)\nonumber\\
&\approx z_{l+1}^{n-\alpha} \left [ 1 + \frac{(n-\alpha)(n-1-\alpha)}{2}\frac{V_{l+1}}{ z_{l+1}^2} \right ],
\label{Eq:happ}
\end{eqnarray}
where we have assumed $\sigma_l/z_l \ll 1$.
Note that \eref{Eq:happ} is exact when $n=\alpha$ or $n=\alpha+1$.

From \eref{Eq:zexact} and \eref{Eq:Vl} along with \eref{Eq:happ}, we get
\begin{eqnarray}
\label{Eq:zlv}
z_{l+1}-z_l &= \frac{a}{\alpha}z_{l+1}^{1-\alpha} 
\left [ 1 + \frac{\alpha(\alpha-1)}{2} \frac{V_{l+1}}{z_{l+1}^2} \right ]
-\dl,\\
V_{l+1}-V_l &= (z_l-\dl)^2 - z_{l+1}^2 
+ \frac{2a}{\alpha}z_{l+1}^{2-\alpha}
\left [ 1 + \frac{(2-\alpha)(1-\alpha)}{2} \frac{V_{l+1}}{z_{l+1}^2} \right ].
\label{Eq:Vlv}
\end{eqnarray}
Using the result \eref{Eq:zlMFA} of the MFA, we see that $V_l$ contributes at best to the subleading behavior 
in \eref{Eq:zlv} if $V_l \approx (C l^{\epsilon})^2 $ with $\epsilon< \gamma/\alpha$. 
The leading behavior of $V_l$, or the value of $\epsilon$, will
be determined from \eref{Eq:Vlv}.

Before finding $\epsilon$, we rewrite \eref{Eq:Vlv} using 
\eref{Eq:zlv} as
\begin{eqnarray}
V_{l+1}-V_l &\approx (z_l-\dl)^2 - z_{l+1}^2 
+ 2z_{l+1} (z_{l+1}-z_l+\dl)
\left [ 1 + (1-\alpha) \frac{V_{l+1}}{z_{l+1}^2} \right ]\nonumber\\
&= (z_{l+1}-z_l+\dl)^2 
+ 2 (1-\alpha) (z_{l+1}-z_l+\dl) \frac{V_{l+1}}{z_{l+1}}\nonumber \\
&=\frac{1}{h(z_{l+1})^2} 
- 2(\alpha-1) \frac{V_{l+1}}{z_{l+1}h(z_{l+1})},
\label{Eq:Vlv2}
\end{eqnarray}
where we have used \eref{Eq:dzdla} and only kept the leading terms.

For  $\alpha \le 1/b$ we have $h(z_l) = \alpha A l /z_l = \alpha (Al)^{1-1/\alpha}/a^{1/\alpha}$, and 
assuming $V_l \approx C^2 l^{2 \epsilon}$ we obtain
\begin{eqnarray}
2 \epsilon C^2 l^{2 \epsilon-1} \doteq \frac{a^{2/\alpha}}{\alpha^2} (Al)^{-2+2/\alpha} - 2 C^2 
\frac{\alpha-1}{\alpha A}l^{2 \epsilon-1},
\end{eqnarray}
which gives
\begin{eqnarray}
\epsilon = \frac{1}{\alpha}-\frac{1}{2},\qquad
C = \frac{(Aa)^{1/\alpha}}{\alpha A \sqrt{ 1 + 2(A^{-1}-1)(1-1/\alpha)}}.
\end{eqnarray}
Since $A=1$ if $\alpha < 1/b$ and $A<1$ if $1/\alpha = b<1$, $C$ is positive.

\begin{figure}[t]
\centering
\includegraphics[width=0.7\textwidth]{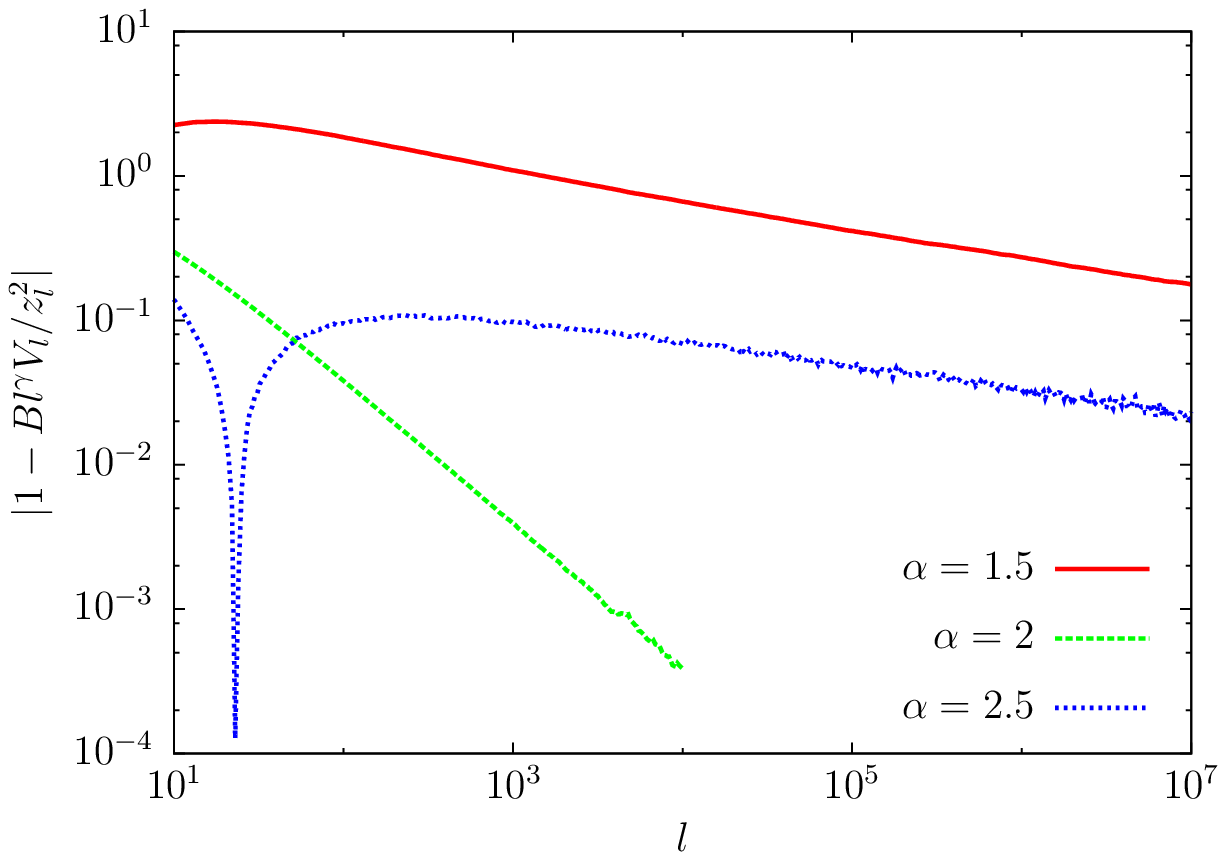}
\caption{\label{Fig:bs_con} Double logarithmic plots of $\left | 1 - B l^\gamma V_l/z_l^2 \right |$
vs $l$ for $\alpha=1.5$, 2, 2.5 with $b=\frac{1}{2}$, $a=c=1$. All curves decay to zero with different powers.
}
\end{figure}
Next we consider the case $\alpha > 1/b$, where $h(z_l)^{-1} = z_l (A l)^{-\gamma} /\alpha = a^{1/\alpha}
(Al)^{(1/\alpha-1)\gamma}/\alpha $, and hence
\begin{eqnarray}
2\epsilon C^2 l^{2\epsilon-1} \doteq
\frac{a^{2/\alpha}(Al)^{2(1/\alpha-1)\gamma}}{\alpha^2} 
- 2 C^2 \frac{\alpha-1}{\alpha A^\gamma} l^{2\epsilon-\gamma}.
\end{eqnarray}
Since $\gamma<1$, the left hand side can at most contribute to the subleading behavior. Thus, we get
\begin{eqnarray}
\epsilon = \frac{\gamma}{\alpha} - \frac{\gamma}{2},\qquad
C = \left [ \frac{(aA^\gamma)^{2/\alpha}}{2 \alpha (\alpha-1) A^\gamma}\right
]^{1/2}.
\end{eqnarray}.

To sum up, we found that
\begin{eqnarray}
\label{Eq:Vlasym}
\frac{V_l}{z_l^2} \approx  \frac{l^{-\gamma}}{B}
\end{eqnarray}
with 
\begin{eqnarray}
B = 
\cases{
\alpha^2 \left [A^2 + 2A(1-A) (1-1/\alpha)\right ],& $\alpha \le 1/b$,\\
2\alpha(\alpha-1) A^\gamma,& $\alpha > 1/b$,
}
\end{eqnarray}
where $\gamma$ and $A$ are given in \eref{Eq:gamma} and \eref{Eq:A}.
Hence, the MFA becomes exact as $l \rightarrow \infty$.

To confirm the above prediction, we performed Monte Carlo simulations 
for various values of $b$, $\alpha$, and $c$. In \fref{Fig:bs_con},
we depict the deviation $|1-B V_l l^{\gamma}/z_l^2|$ from \eref{Eq:Vlasym} against $l$ for 
$b=\frac{1}{2}$, $c=1$, and  $\alpha = 1.5,2,2.5$ on a double-logarithmic scale. 
As predicted, all curves approach zero.

\section*{References}
\bibliographystyle{iopart-num}
\bibliography{me.bib}
\end{document}